\documentclass[superscriptaddress,onecolumn,pre]{revtex4}

\usepackage{amsmath}
\usepackage{graphicx,color}

\newcommand{\be}{\begin{equation}}
\newcommand{\ee}{\end{equation}}
\newcommand{\bea}{\begin{eqnarray}}
\newcommand{\eea}{\end{eqnarray}}

\begin{document}
\sloppy

%-title page-%

\title{Kinetic theory of Onsager's vortices in two-dimensional hydrodynamics}

\author{Pierre-Henri Chavanis}
%\email{chavanis@irsamc.ups-tlse.fr}
\affiliation{Laboratoire de Physique Th\'eorique (IRSAMC), CNRS and UPS, Universit\'e de Toulouse, F-31062 Toulouse, France}

\begin{abstract}
Starting from the Liouville equation, and using a BBGKY-like hierarchy, we derive a kinetic equation for the point vortex gas in two-dimensional (2D) hydrodynamics, taking two-body correlations and collective effects into account. This equation is valid at the order $1/N$ where $N\gg 1$ is the number of point vortices in the system (we assume that their individual circulation scales like $\gamma\sim 1/N$). It gives the first correction, due to graininess and correlation effects, to the 2D Euler equation that is obtained for $N\rightarrow +\infty$. For axisymmetric distributions, this kinetic equation does {\it not} relax towards the Boltzmann distribution of statistical equilibrium. This implies either that (i) the ``collisional'' (correlational) relaxation time is larger than $Nt_D$, where $t_D$ is the dynamical time, so that three-body, four-body... correlations must be taken into account in the kinetic theory, or (ii) that the point vortex gas is non-ergodic (or does not mix well) and will never attain statistical equilibrium. Non-axisymmetric distributions may relax towards the Boltzmann distribution on a timescale of the order $Nt_D$ due to the existence of additional resonances, but this is hard to prove from the kinetic theory. On the other hand, 2D Euler unstable vortex distributions can experience a process of ``collisionless'' (correlationless) violent relaxation towards a non-Boltzmannian quasistationary state (QSS) on a very short timescale of the order of a few dynamical times. This QSS is possibly described by the Miller-Robert-Sommeria (MRS) statistical theory which is the counterpart, in the context of two-dimensional hydrodynamics, of the Lynden-Bell statistical theory of violent relaxation in stellar dynamics.
\end{abstract}

\maketitle

\section{Introduction}
\label{sec_introduction}

In 1949, Onsager \cite{onsager} published a seminal paper in which he
laid down the foundations of the statistical mechanics of vortices in
two-dimensional hydrodynamics \footnote{His paper contains two
additional major results: (i) the quantization of the circulation of
vortices in superfluids (this is mentioned as a footnote in his
paper); (ii) the inviscid dissipation of energy in 3D turbulence for
singular Euler solutions.}. He considered the point vortex gas as an
idealization of more realistic vorticity fields and discovered that
negative temperature states are possible for this system \footnote{His
argument for the existence of negative temperatures was given two
years before Purcell \& Pound
\cite{purcell} reported the presence of negative ``spin temperatures''
in an experiment on nuclear spin systems.}. At negative temperatures, corresponding to high energies,
like-sign vortices have the tendency to cluster into
``supervortices'' similar to the large-scale vortices (e.g. Jupiter's
great red spot) observed in the atmosphere of giant planets. If the point vortices all have the same sign, one gets a monopole. If they have different signs, one gets a dipole made of two clusters of opposite sign, or possibly a tripole made of a central vortex of a given sign surrounded by two vortices of opposite sign. Therefore, statistical mechanics may explain the ubiquity of large-scale vortices observed in geophysical and astrophysical flows.

The qualitative arguments of Onsager were developed more
quantitatively in a mean field approximation by Joyce \& Montgomery
\cite{jm,mj}, Kida \cite{kida} and Pointin \& Lundgren \cite{pl,lp}, and by Onsager himself in
unpublished notes \cite{esree}. The statistical theory predicts that
the point vortex gas should relax towards an equilibrium state
described by the Boltzmann distribution.  Specifically, the
equilibrium stream function is solution of a Boltzmann-Poisson
equation. Several mathematical works \cite{caglioti,k93,es,ca2,kl}
have shown how a proper thermodynamic limit could be rigorously
defined for the point vortex gas (in the Onsager picture). It is shown
that the mean field approximation becomes exact in the limit
$N\rightarrow +\infty$ with $\gamma\sim 1/N$ (where $N$ is the number
of point vortices and $\gamma$ is the circulation of a point vortex).

A practical limitation of Onsager's theory resides in the
approximation of a continuous flow by a discrete collection of point
vortices. This is clearly an idealization that he was aware of: ``The
distributions of vorticity which occur in the actual flow of normal
liquids are continuous''
\cite{onsager}. This approximation also leads to some ambiguity in the
construction of a statistical model of realistic flows: ``What set of
discrete vortices will best approximate a continuous distribution of
vorticity?''
\cite{onsager}.

The statistical mechanics of continuous vorticity fields was developed
later  by Miller \cite{miller} and Robert \& Sommeria
\cite{rs}. The same theory appears in a  paper (in russian) by Kuzmin \cite{kuzmin} eight years before but his
contribution is less well-known. The Miller-Robert-Sommeria (MRS)
statistical theory is based on the 2D Euler equation which describes
incompressible and inviscid flows. It predicts that the 2D Euler
equation can reach a statistical equilibrium state (or metaequilibrium
state) on a coarse-grained scale as a result of a mixing process.
Recently, the MRS theory has been applied to geophysical and
astrophysical flows, notably to oceanic circulation \cite{kazantsev},
jovian vortices (Jupiter's great red spot) \cite{turkington,bs,sw},
Fofonoff flows
\cite{vb,vblong,naso} and oceanic rings and jets \cite{vr}.

The MRS theory shares many analogies with the theory of violent
relaxation developed by Lynden-Bell \cite{lb} for collisionless
stellar systems described by the Vlasov equation.  The analogy between
two-dimensional vortices and self-gravitating systems was mentioned by
Onsager in a letter to Lin: ``At {\it negative} temperatures, the
appropriate statistical methods have analogues not in the theory of
electrolytes, but in the statistics of stars...'' \cite{esree}. This
analogy has been further developed in \cite{houchesPH}.

At that stage, we would like to make some observations that will be
important in the following.  First of all, point vortices do not exist
in nature \footnote{An exception concerns the case of non-neutral
plasmas under a strong magnetic field \cite{jm,mj}. In the so-called
``guiding-center approximation'', this system consists in long
filaments of charge parallel to the magnetic field ${\bf B}$ and
moving under their mutual electric field ${\bf E}$ with the ${\bf
E}\times {\bf B}/B^2$ velocity. The mathematical description of this
system is equivalent to a collection of $N$ point vortices in which
the charge of a filament corresponds to the circulation of a
vortex.}. Fundamentally, the physical system to consider is the
continuous vorticity field that is solution of the 2D Euler
equation. The point vortex model is a particular solution of the 2D
Euler equation of the form $\omega({\bf
r},t)=\sum_{i=1}^N\gamma_i\delta({\bf r}-{\bf r}_i(t))$. However,
there are rigorous results
\cite{mp} which show that any smooth solution $\omega({\bf r},t)$ of
the 2D Euler equation may be approximated arbitrarily well over a
finite time interval by a collection of point vortices with
$\gamma_i\sim 1/N$ as $N\rightarrow +\infty$. The physical situation
is completely different in the case of stellar systems. Basically, a
stellar system is a discrete collection of $N$ point mass stars. For
$N\rightarrow +\infty$ in a proper thermodynamic limit, this discrete
system can be approximated by a continuous distribution function that
is solution of the Vlasov equation. This is the reverse situation with
respect to 2D hydrodynamics! It is nevertheless interesting to further
develop the analogy between a system of $N$ stars in astrophysics and
a system of $N$ point vortices in 2D hydrodynamics.

It is well-known in astrophysics that a stellar system undergoes two
successive types of relaxation \cite{henon}. In the ``collisionless
regime'', one can ignore correlations between stars due to finite $N$
effects and make a mean field approximation. In that case, the
evolution of the system is governed by the Vlasov equation. The
Vlasov-Poisson system can undergo a violent collisionless relaxation
towards a quasi stationary state (QSS) on a few dynamical times. This
corresponds to the mixing of the distribution function by the mean
field. This mixing process is responsible for the apparent regularity
of galaxies. This is precisely what the Lynden-Bell \cite{lb}
statistical theory tries to describe. Unfortunatelly, in astrophysics,
the Lynden-Bell theory does not give a good prediction of the
structure of the whole galaxy because of the problem of {\it
incomplete relaxation} in the outer part of the system where mixing is
inefficient.  On a longer timescale, in the so-called ``collisional
regime'', one must take into account correlations between stars due to
finite $N$ effects (graininess)
\footnote{Actually, the evolution of the distribution function is not due to
real collisions between stars (like in the theory of gases) but to
weak deflexions due to binary encounters with relatively large impact
parameter \cite{bt}. For simplicity, we shall however call them
``collisions''.}. These correlations are expected to drive the system
towards the ordinary statistical equilibrium state described by the
Boltzmann distribution for $t\rightarrow +\infty$.  This mixing
process is due to discrete effects and takes place on a very long
timescale. Indeed, the ``collisional'' relaxation time scales like
$(N/\ln N)t_D$ and it diverges for $N\rightarrow +\infty$
\cite{bt}. For self-gravitating systems, the Boltzmann distribution is
not reached in practice because of evaporation \cite{spitzer} and the
gravothermal catastrophe
\cite{antonov,lbw}.  Even if the statistical mechanics of
self-gravitating systems is complicated because of the peculiar
nature of the $1/r$ interaction, the concepts of violent collisionless
relaxation towards a QSS and of slow collisional relaxation towards
the Boltzmann distribution are fundamental and can be transposed to
other systems with long-range interactions. For example, they have
been clearly illustrated for the HMF model \cite{cdr}.

The same distinction between collisionless and collisional relaxation
applies to the system of point vortices. In the ``collisionless
regime'', one can ignore correlations between point vortices due to
finite $N$ effects and make a mean field approximation. In that case,
the evolution of the system is governed by the 2D Euler equation. The
2D Euler-Poisson system can undergo a process of violent collisionless
relaxation towards a quasi stationary state (QSS) on a few dynamical
times.  This corresponds to the mixing of the vorticity by the mean
field.  This mixing process leads to the formation of large scale
vortices similar to those seen in planetary atmospheres.  This is
precisely what the MRS statistical theory tries to describe.  On a
longer timescale, called the ``collisional regime'', one must take
into account correlations between point vortices due to finite $N$
effects (graininess)
\footnote{Here again, the evolution of the vorticity field is not
due to real collisions between point vortices but to distant
binary encounters. For simplicity, we shall however call them
``collisions''.}. These correlations are expected to drive the system
towards the ordinary statistical equilibrium state, described by the
Boltzmann distribution, for $t\rightarrow +\infty$.  This mixing
process is due to discrete effects and takes place on a very long
timescale. As we shall see, the scaling with $N$ of the
``collisional'' relaxation time of point vortices is not firmly
established.

It is of paramount importance not to confuse the Lynden-Bell (or MRS)
statistical theory and the Boltzmann statistical theory which apply to
very different timescales in the evolution of a Hamiltonian system
with long-range interactions. This is related to the non-commutation
of the limits $N\rightarrow +\infty$ and $t\rightarrow +\infty$.
Mathematically speaking, the QSS is reached when the $N\rightarrow
+\infty$ limit is taken before the $t\rightarrow +\infty$ limit while
the Boltzmann distribution is reached when the $t\rightarrow +\infty$
limit is taken before the $N\rightarrow +\infty$ limit. The
Lynden-Bell (or MRS) theory makes the statistical mechanics of a {\it
continuous field} evolving according to the Vlasov (or 2D Euler)
equation while the Boltzmann theory makes the statistical mechanics of
a {\it discrete system of particles} (stars or point vortices)
evolving according to Hamiltonian equations. The first describes the
QSS that is formed after a few dynamical times and the second
describes the statistical equilibrium state that is reached for
$t\rightarrow +\infty$. The distribution predicted by Lynden-Bell (or
MRS) is different from the Boltzmann distribution due to additional
conservation laws in the Vlasov (or Euler) equation. In fact, the
Lynden-Bell (or MRS) statistical mechanics and the Boltzmann
statistical mechanics describe two types of mixing occurring at
different scales: in the process of violent ``collisionless''
relaxation, the smooth distribution function $f({\bf r},{\bf v},t)$
(or vorticity field $\omega({\bf r},t)$) that evolves according to the
Vlasov (or Euler) equation is mixed by the mean field and the {\it
coarse-grained} distribution function $\overline{f}({\bf r},{\bf
v},t)$ (or vorticity field $\overline{\omega}({\bf r},t)$) reaches a
statistical equilibrium state. In the process of ``collisional''
relaxation, the discrete distribution of particles $f_d({\bf r},{\bf
v},t)=\sum_i m \delta({\bf r}-{\bf r}_i(t))\delta({\bf v}-{\bf
v}_i(t))$ or $\omega_d({\bf r},t)=\sum_i \gamma \delta({\bf r}-{\bf
r}_i(t))$ (stars or point vortices) that evolves according to the
Klimontovich equation (equivalent to the Hamilton equations) is mixed
by discrete effects and the {\it smooth} distribution of particles
${f}({\bf r},{\bf v},t)$ or ${\omega}({\bf r},t)$ reaches a
statistical equilibrium state.

Onsager \cite{onsager} assumed ergodicity of the point vortex gas and
determined the ordinary statistical equilibrium state by evaluating
the density of states and the entropy of the Hamiltonian
system. However, he did not mention the source of mixing leading to
the statistical equilibrium state. As we have explained, there are two
sources of mixing in the point vortex gas: the fluctuations of the
mean field during the phase of violent relaxation and the fluctuations
due to discrete effects during the phase of collisional
relaxation. The first one is very rapid and the second one very
slow. It is likely that Onsager had in mind the mixing due to the
fluctuations of the mean field. Indeed, he was basically interested in
the 2D Euler equation, not by finite $N$ effects (``collisions'') in
the point vortex gas. As we have explained, only continuous vorticity
fields make sense in 2D hydrodynamics. The fact that he used the point
vortex model was just a question of commodity. At that time, nobody
knew how to make the statistical mechanics of a continuous vorticity
flow. However, the statistical equilibrium state of the 2D Euler
equation should be described by the MRS distribution, not by the
Boltzmann distribution. The Boltzmann distribution describes the
statistical equilibrium state of the point vortex gas that is reached
for $t\rightarrow +\infty$ due to the development of correlations
(finite $N$ effects) between point vortices. It is not clear whether
Onsager had anticipated these two very different regimes related to
the subtle non-commutation of the $t\rightarrow +\infty$ and
$N\rightarrow +\infty$ limits. 

In the present paper, we shall be essentially interested in the slow
``collisional'' relaxation of the point vortex gas towards the
Boltzmann distribution due to finite $N$ effects. Its fast
``collisionless'' relaxation towards a QSS for $N\rightarrow +\infty$
will be only briefly discussed. Now, the relevance of the Boltzmann
distribution for point vortices when $t\rightarrow +\infty$ is not
clearly established because it relies on an assumption of ergodicity
(or sufficient mixing) that may not be realized \footnote{In fact, it
has been proven that the point vortex gas is non-ergodic
\cite{khanin}. However, statistical mechanics does not require {\it
strict} ergodicity. An efficient mixing over the energy surface will
suffice to justify the use of the microcanonical ensemble and of the
Boltzmann distribution for $N\rightarrow +\infty$. However, our
concern here is that the point vortex gas may not mix well enough due
to the effect of ``collisions'' (correlations, finite $N$ effects) to
justify the applicability of statistical mechanics and of the
Boltzmann distribution.}.  If we want to prove that the point vortex
gas truly relaxes towards the Boltzmann distribution for $t\rightarrow
+\infty$, and if we want to determine the relaxation time (in
particular its scaling with $N$), we must develop a kinetic
theory. The MRS statistical theory is also based on an assumption of
ergodicity, or efficient mixing, and the process of violent relaxation
of the 2D Euler equation towards a QSS should also be described in
terms of a kinetic theory (for the coarse-grained vorticity
field). This kinetic theory is more complicated to develop and will
not be considered here. We only refer to
\cite{rsmepp,rr,csr,quasi} for some investigations in this direction.

In previous papers \cite{preR,pre,bbgky,kindetail}, we have developed a
kinetic theory of 2D point vortices and we have derived a kinetic
equation for the evolution of the smooth vorticity field taking
two-body correlations into account. This equation is valid at the
order $1/N$ and provides therefore the first order correction to the 
Euler equation obtained for $N\rightarrow +\infty$. This kinetic
equation was derived by using various methods such as the projection
operator formalism, the quasilinear theory and the BBGKY hierarchy. In
these works, we focused on (distant) two-body collisions and neglected
collective effects. This leads to a kinetic equation similar to the
Landau \cite{landau} equation in plasma physics \footnote{Contrary to
the Landau equation, our derived kinetic equation does not exhibit any
divergence, suggesting that collective effects are not crucial for
point vortices contrary to the case of plasma physics where they
account for Debye shielding and regularize the large-scale divergence
appearing in the Landau equation.}. A kinetic theory of the point
vortex gas had been previously developed by Dubin \& O'Neil
\cite{dubin} in the context of non-neutral plasmas under a strong
magnetic field. They used the Klimontovich formalism and took
collective effects into account. This leads to a kinetic equation
similar to the Lenard-Balescu \cite{lenard,balescu} equation in plasma
physics. Their work was pursued in
\cite{sdprl,sd2,dubinjin,dubin2} in different directions. In a recent paper
\cite{klim}, we used the Klimontovich formalism to derive a
Fokker-Planck equation describing the evolution of a test vortex in a
bath of field vortices, taking collective effects into account.

The purpose of the present paper is two-fold. One, physical, motivation  is to discuss the implications of the kinetic theory of two-dimensional point vortices. An important conclusion of our analysis is that  the relevance of the Boltzmann distribution is not established by the present-day kinetic theory. More precisely, we show that, if the point vortex gas ever relaxes towards the statistical equilibrium state, this takes place on a very long time, larger than $Nt_D$ (for axisymmetric distributions). The collisional relaxation of point vortices is therefore a very slow process and requires to take into account three-body, four-body... correlations, of order $1/N^2$, $1/N^3$..., that are ignored so far in the kinetic theory.    It is also possible that mixing by ``collisions'' is not sufficient to drive the system towards the Boltzmann distribution. We emphasize, however, that the point vortex gas can rapidly reach a QSS as a result of a ``collisionless'' violent relaxation. However, this QSS is described by the Miller-Robert-Sommeria, or Lynden-Bell, distribution, not by the Boltzmann distribution.  Another, more technical, motivation of our paper is to derive  the Lenard-Balescu-type kinetic equation for point vortices from the BBGKY hierarchy. This derivation (which constitutes the main result of the paper) is new and completes our previous derivation \cite{bbgky} where collective effects were neglected.  Our approach closely follows the method of Ichimaru \cite{ichimaru}  for deriving the Lenard-Balescu equation in plasma physics.  We also consider the relaxation of a test vortex in a bath of field vortices described by a Fokker-Planck equation. Finally, we mention open problems and future directions of research.

We note that other kinetic theories of point vortices 
exist but they apply to situations different from the one we will
consider here. For example, Nazarenko \& Zakharov \cite{zakharov}
obtained a kinetic equation for point vortices with different
intensities moving on the background of a shear flow and experiencing
``hard'' collisions. Marmanis \cite{marmanis} and Newton
\& Mezic \cite{newtonmezic} derived kinetic equations for a vortex gas
viewed as a coupling, via the Liouville equation, between monopoles,
dipoles and tripoles. Finally, Sire \& Chavanis \cite{renormalization}
developed a kinetic theory of three-body collisions (dipoles hitting
monopoles) with application to the context of 2D decaying
turbulence. General results about point vortices can be found in the
book of Newton \cite{newton}.

\section{Two-dimensional point vortices: Evolution of the system as a whole}

\subsection{The Boltzmann distribution}
\label{sec_nv}

We consider an isolated system of point vortices with identical
circulation $\gamma$ on an infinite plane. Their dynamics is fully
described by the Kirchhoff-Hamilton equations \cite{kirchhoff,newton}:
\begin{eqnarray}
\label{nv1} \gamma{d{x}_{i}\over dt}={\partial H\over\partial {y}_{i}},
\qquad \gamma{d{y}_{i}\over dt}=-{\partial H\over\partial {x}_{i}}, \qquad
H=-\frac{\gamma^{2}}{2\pi}\sum_{i<j}\ln |{\bf r}_{i}-{\bf r}_{j}|,
\end{eqnarray}
where the coordinates $(x,y)$ of the point vortices are canonically
conjugate. We shall denote the potential of interaction by $u({\bf r},{\bf r}')=-\frac{1}{2\pi}\ln |{\bf r}-{\bf r}'|$. This Hamiltonian system conserves the energy $E=H$, the
circulation $\Gamma=N\gamma$, the angular momentum $L=\sum_{i} \gamma
r_{i}^{2}$ and the impulse ${\bf P}=\sum_{i}\gamma {\bf r}_{i}$. We
take the origin of the system of coordinates at the center of vorticity so that ${\bf P}={\bf 0}$
(we shall ignore this constraint in the following).  The angular momentum fixes the vortex size $R$. For $t\rightarrow
+\infty$, we expect this system to reach a statistical equilibrium
state described by the microcanonical distribution
\cite{onsager,lp,bbgky}:
\begin{equation}
\label{nv2} P_{N}({\bf r}_{1},...,{\bf r}_{N})={1\over
g(E,L)} \delta(E-H({\bf r}_{1},...,{\bf r}_{N}))\delta(L-\sum_{i}\gamma
r_{i}^{2}),
\end{equation}
where $g(E,L)=\int \delta(E-H)\delta(L-\sum_{i}\gamma
r_{i}^{2})\prod_{i}d{\bf r}_{i}$ is the density of states. The
microcanonical entropy of the system is defined by $S(E,L)=\ln
g(E,L)$.  The microcanonical temperature and the angular velocity are
then given by $\beta=1/T=(\partial S/\partial E)_{L}$ and
$\Omega_{L}=(2/\beta)(\partial S/\partial L)_{E}$. As first realized
by Onsager \cite{onsager}, the temperature of the point vortex gas can
be negative.  At negative temperatures $\beta(E)<0$, corresponding to
high energy states, point vortices of the same sign group themselves
in ``supervortices'' similar to large-scale vortices in planetary atmospheres. 
We define the thermodynamic limit as $N\rightarrow
+\infty$ in such a way that the normalized energy
$\epsilon=E/\Gamma^2$, the normalized temperature $\eta=\beta
\Gamma \gamma$, the normalized angular momentum $\lambda=L/\Gamma R^2$ and the normalized angular velocity $v=\Omega_L R^2/\Gamma$ are of order unity (these scalings result from simple dimensional analysis).
We can renormalize the parameters so that the circulation of the
vortices scales like $\gamma\sim 1/N$ and the vortex size like $R\sim
1$. Then, the inverse temperature scales like $\beta\sim N$, the
energy like $E\sim 1$, the angular momentum  like $L\sim 1$ and the angular velocity like $\Omega_L\sim 1$. On the other
hand, the total circulation $\Gamma\sim N\gamma\sim 1$ and the
dynamical time $t_{D}\sim 1/\omega\sim R^2/\Gamma\sim 1$ are of order
unity.

In the thermodynamic limit $N\rightarrow +\infty$ defined previously,
it can be rigorously proven \cite{caglioti,k93,es,kl,ca2} that the $N$-body
distribution function factorizes in a product of $N$ one-body
distribution functions
\begin{equation}
\label{prod} P_{N}({\bf r}_{1},...,{\bf
r}_{N})=\prod_{i=1}^{N} P_{1}({\bf r}_{i}).
\end{equation}
Therefore, the mean
field approximation becomes exact in the thermodynamic limit
$N\rightarrow +\infty$.  Furthermore, the one-body distribution
function $P_{1}({\bf r})$, or equivalently the smooth density of point
vortices $n({\bf r})=N P_{1}({\bf r})$ or the smooth vorticity field
$\omega({\bf r})=N\gamma P_{1}({\bf r})$, is the solution of a maximum entropy principle \cite{mj,houchesPH}:
\begin{eqnarray}
\label{nv3}
S(E,\Gamma,L)=\max_{\omega}\quad \lbrace S_{B}[{\omega}]\quad |\quad E[{\omega}]=E, \ \Gamma[{\omega}]=\Gamma, \ L[{\omega}]=L \rbrace,
\end{eqnarray}
where
\begin{equation}
\label{nv4} S_{B}[\omega]=-\int {\omega\over \gamma}\ln {\omega\over \gamma} \, d{\bf r}, \qquad E={1\over 2}\int  \omega\psi \, d{\bf r},\qquad \Gamma=\int \omega \, d{\bf r},\qquad L=\int \omega r^2\, d{\bf r},
\end{equation}
are the Boltzmann entropy, the mean field energy, the circulation and
the angular momentum. Here, $\psi$ is  the mean field stream function produced by the smooth distribution of vortices $\omega$ according to the Poisson equation
\begin{equation}
\label{poissonv}
-\Delta\psi=\omega.
\end{equation}
The mean velocity of a point vortex is then $\langle{\bf
V}\rangle=-{\bf z}\times \nabla\psi$ where ${\bf z}$ is a unit vector
normal to the flow. Fundamentally, the Boltzmann entropy is defined by $S_B=\ln W$ where $W$ is the number of {\it microstates} (complexions), specified by the precise position $\lbrace {\bf r}_1,...,{\bf r}_N\rbrace$ of each point vortex, corresponding to a given {\it macrostate}, specified by the smooth vorticity
field $\lbrace \omega({\bf r})\rbrace$ giving the average number of point vortices in macrocells of size $0<\Delta\ll 1$. Using the Stirling formula for $N\gg 1$, we obtain the expression (\ref{nv4}) of the Boltzmann entropy. Introducing Lagrange multipliers and writing
the variational principle in the form $\delta S_{B}-\beta\delta
E-\alpha\delta \Gamma-\beta \frac{\Omega_{L}}{2}\delta L=0$, we obtain
the mean field Boltzmann distribution
\begin{equation}
\label{nv5} \omega=A\gamma e^{-\beta \gamma (\psi+\frac{\Omega_{L}}{2}r^{2})},
\end{equation}
where $A$ is a positive constant.  Substituting this relation in the Poisson
equation (\ref{poissonv}), we obtain the Boltzmann-Poisson
equation
\begin{equation}
\label{nv6} -\Delta\psi=A\gamma e^{-\beta \gamma(\psi+\frac{\Omega_{L}}{2}r^{2})},
\end{equation}
like in the theory of electrolytes (for $\beta>0$) or in the
statistics of stars (for $\beta<0$). The statistical equilibrium state
is then obtained by solving this equation and relating the Lagrange
multipliers $\beta$, $\alpha$ (or $A$) and $\Omega_{L}$ to the
constraints $E$, $\Gamma$ and $L$ \cite{lp}. Then, we have to make
sure that the resulting distribution is a {\it maximum} of $S_{B}$ at
fixed circulation, energy and angular momentum (most probable state),
not a minimum or a saddle point.

%In an infinite domain, the statistical
%equilibrium state is always axisymmetric \cite{lp}. In a bounded
%domain, the statistical equilibrium state crucially depends on the
%domain shape and several situations have been considered, see
%e.g. \cite{jm,lp,smith,caglioti,houchesPH}.

\subsection{BBGKY-like hierarchy and $1/N$ expansion}
\label{sec_bv}

In order to establish whether the point vortex gas will reach the Boltzmann
distribution (\ref{nv5}) predicted by statistical mechanics and
determine the timescale of the relaxation, in particular its scaling with $N$, we need to develop a
kinetic theory of point vortices. Basically, the evolution of the $N$-body distribution $P_N({\bf r}_1,...,{\bf r}_N,t)$
of the point vortex gas is governed by the Liouville equation
\begin{eqnarray}
\label{bv1}
\frac{\partial P_N}{\partial t}+\sum_{i=1}^{N}{\bf
V}_{i}\cdot \frac{\partial P_N}{\partial {\bf r}_i}=0,
\end{eqnarray}
where
\begin{eqnarray}
\label{bv2}
{\bf V}_{i}=-{\bf z}\times \frac{\partial\psi}{\partial {\bf r}_{i}}({\bf r}_i)=\frac{\gamma}{2\pi}{\bf z}\times \sum_{j\neq i}\frac{{\bf r}_{i}-{\bf r}_{j}}{|{\bf r}_{i}-{\bf r}_{j}|^{2}}=\sum_{j\neq i}{\bf V}(j\rightarrow i),
\end{eqnarray}
is the velocity of point vortex $i$ due to its interaction with the
other vortices. Here, $\psi({\bf r}_i)$ denotes the exact stream function in ${\bf r}_i$
produced by the discrete distribution of point vortices and ${\bf
V}(j\rightarrow i)$ denotes the exact velocity induced by point vortex $j$
on point vortex $i$. The Liouville equation (\ref{bv1}), which  is equivalent to the
Hamiltonian system (\ref{nv1}), contains
too much information to be of practical use.
In general, we are only
interested in the one-body or two-body distributions \footnote{Nevertheless, a conclusion of our study will be that higher order distributions should also be considered.}. From the Liouville equation
(\ref{bv1}) we can construct a complete BBGKY-like hierarchy for the
reduced distributions $P_{j}({\bf r}_{1},...,{\bf
r}_{j},t)=\int P_{N}({\bf r}_{1},...,{\bf r}_{N},t)\, d{\bf
r}_{j+1}...d{\bf r}_{N}$. It reads \cite{bbgky}:
\begin{eqnarray}
\frac{\partial P_j}{\partial t}+\sum_{i=1}^{j}\sum_{k=1,k\neq
i}^{j}{\bf V}(k\rightarrow i)\cdot \frac{\partial P_j}{\partial {\bf
r}_i}+(N-j)\sum_{i=1}^{j}\int {\bf V}(j+1\rightarrow
i)\cdot \frac{\partial P_{j+1}}{\partial {\bf r}_i}\, d{\bf r}_{j+1}=0.
 \label{bv4}
\end{eqnarray}
This hierarchy of equations is not closed since the equation for the
one-body distribution $P_{1}({\bf r}_{1},t)$ involves the two-body
distribution $P_{2}({\bf r}_{1},{\bf r}_{2},t)$, the equation for the
two-body distribution $P_{2}({\bf r}_{1},{\bf r}_{2},t)$ involves the
three-body distribution $P_{3}({\bf r}_{1},{\bf r}_{2},{\bf
r}_{3},t)$, and so on... The idea is to close the hierarchy by using a
systematic expansion of the solutions in powers of $1/N$ in the
thermodynamic limit $N\rightarrow +\infty$.

The first two equations of the hierarchy are
\begin{eqnarray}
\frac{\partial P_1}{\partial t}({\bf r}_{1},t)+(N-1)\int {\bf V}(2\rightarrow
1)\cdot \frac{\partial P_{2}}{\partial {\bf r}_1}({\bf r}_{1},{\bf r}_{2},t)\, d{\bf r}_{2}=0.
 \label{bv4a}
\end{eqnarray}
\begin{eqnarray}
\frac{1}{2}\frac{\partial P_2}{\partial t}({\bf r}_{1},{\bf r}_{2},t)+{\bf V}(2\rightarrow 1)\cdot \frac{\partial P_2}{\partial {\bf
r}_1}({\bf r}_{1},{\bf r}_{2},t)+(N-2)\int {\bf V}(3\rightarrow
1)\cdot \frac{\partial P_{3}}{\partial {\bf r}_1}({\bf r}_{1},{\bf r}_{2},{\bf r}_{3},t)\, d{\bf r}_{3}+(1\leftrightarrow 2)=0.
 \label{bv4b}
\end{eqnarray}
We decompose the two- and three-body distributions in the
suggestive form
\begin{equation}
\label{bv5}
P_{2}({\bf r}_{1},{\bf r}_{2},t)=P_{1}({\bf r}_{1},t)P_{1}({\bf r}_{2},t)+P_{2}'({\bf r}_{1},{\bf r}_{2},t),
\end{equation}
\begin{eqnarray}
\label{bv6}
P_{3}({\bf r}_{1},{\bf r}_{2},{\bf r}_{3},t)=P_{1}({\bf r}_{1},t)P_{1}({\bf r}_{2},t)P_{1}({\bf r}_{3},t)+P_{2}'({\bf r}_{1},{\bf r}_{2},t)P_{1}({\bf r}_{3},t)\nonumber\\
+P_{2}'({\bf r}_{1},{\bf r}_{3},t)P_{1}({\bf r}_{2},t)+P_{2}'({\bf r}_{2},{\bf r}_{3},t)P_{1}({\bf r}_{1},t)+P_{3}'({\bf r}_{1},{\bf r}_{2},{\bf r}_{3},t).
\end{eqnarray}
This is similar to the Mayer \cite{mayer} decomposition in plasma physics. Substituting Eqs. (\ref{bv5}) and (\ref{bv6}) in Eqs. (\ref{bv4a}) and (\ref{bv4b}) and simplifying some terms, we obtain
\begin{eqnarray}
\frac{\partial P_1}{\partial t}({\bf r}_1,t)+(N-1)\left\lbrack \int {\bf V}(2\rightarrow
1)P_{1}({\bf r}_{2},t)\, d{\bf r}_{2}\right\rbrack\cdot \frac{\partial P_1}{\partial {\bf r}_{1}}({\bf r}_1,t) =-(N-1)\frac{\partial}{\partial {\bf r}_1}\cdot \int {\bf V}(2\rightarrow
1) P'_{2}({\bf r}_{1},{\bf r}_{2},t)\, d{\bf r}_{2},
 \label{bv6a}
\end{eqnarray}
\begin{eqnarray}
\frac{1}{2}\frac{\partial P'_2}{\partial t}({\bf r}_{1},{\bf r}_{2},t)
+(N-2) \left\lbrack \int {\bf V}(3\rightarrow
1) P_1({\bf r}_3,t)\, d{\bf r}_{3} \right\rbrack \cdot \frac{\partial P'_{2}}{\partial {\bf r}_1}({\bf r}_{1},{\bf r}_{2},t)\nonumber\\
+\left\lbrack {\bf V}(2\rightarrow 1)- \int {\bf V}(3\rightarrow
1) P_1({\bf r}_{3},t)\, d{\bf r}_{3}\right\rbrack \cdot \frac{\partial P_1}{\partial {\bf
r}_1}({\bf r}_{1},t)P_1({\bf r}_2,t)+{\bf V}(2\rightarrow 1)\cdot \frac{\partial P'_2}{\partial {\bf
r}_1}({\bf r}_{1},{\bf r}_{2},t)\nonumber\\
+(N-2)\left\lbrack \int {\bf V}(3\rightarrow
1)  P'_{2}({\bf r}_{2},{\bf r}_{3},t)\, d{\bf r}_{3}\right\rbrack \cdot \frac{\partial P_1}{\partial {\bf r}_1}({\bf r}_1,t)
\nonumber\\
-\frac{\partial}{\partial {\bf r}_1}\cdot \int {\bf V}(3\rightarrow
1)  P'_{2}({\bf r}_{1},{\bf r}_{3},t)P_1({\bf r}_2,t)\, d{\bf r}_{3}
+(N-2)\frac{\partial}{\partial {\bf r}_1}\cdot \int {\bf V}(3\rightarrow
1)  P'_{3}({\bf r}_{1},{\bf r}_{2},{\bf r}_{3},t)\, d{\bf r}_{3}+(1\leftrightarrow 2)=0.
 \label{bv6b}
\end{eqnarray}
Equations (\ref{bv6a}) and (\ref{bv6b}) are exact for all $N$ but they are not closed. We shall close these equations by expanding the solutions in powers
of $1/N$ for $N\gg 1$. In this limit, the correlation functions
$P'_{j}({\bf r}_{1},...,{\bf r}_{j},t)$ scale like $1/N^{j-1}$. In particular, $P'_2\sim 1/N$ and $P'_3\sim 1/N^2$. On the other hand, $P_1\sim 1$ and $|{\bf V}(i\rightarrow j)|\sim \gamma\sim 1/N$. We are aiming at obtaining a kinetic equation that is valid at the order $1/N$. Let us consider the terms in Eq. (\ref{bv6b}) one by one. The first and second terms are of order $1/N$. They represent the transport of the two-body correlation function by the mean flow.  The third term represents the effect of ``soft'' binary  collisions between vortices; it is of order $1/N$. If we consider only these first three terms (as done in our previous paper \cite{bbgky}), we obtain a kinetic equation that is the counterpart of the Landau equation in plasma physics. The fourth term represents the effect of ``hard'' binary collisions between vortices. This is the term which, together with the previous ones, gives rise to the Boltzmann equation in the theory of gases. It is of order $1/N^2$ but it may become large at small scales so its effect is not entirely negligible. For example, in plasma physics, hard collisions must be taken into account in order to regularize the logarithmic divergence that appears at small scales in the Landau and Lenard-Balescu equations. In the case of point vortices, there is no divergence at small scales in the kinetic equation that we shall obtain. Therefore, in this paper, we shall ignore the contribution of this term (but we note that it would be interesting to study it specifically). The fifth term is of order $1/N$ and it corresponds to collective effects. In plasma physics, this term leads to the Lenard-Balescu equation. It takes into account dynamical Debye screening and regularizes the divergence at large scales that appears in the Landau equation. The main contribution of this work will be to take this term into account in the kinetic theory of point vortices in order to obtain a Lenard-Balescu-type kinetic equation from the BBGKY hierarchy.  The last two terms are of the order $1/N^2$ and they will be neglected. In particular, at the order $1/N$, we can neglect the three-body correlation function. In this way, the hierarchy of equations is closed.

If we introduce the smooth vorticity field
$\omega({\bf r}_1,t)=N\gamma P_{1}({\bf r}_1,t)$  and the two-body correlation function $g({\bf r}_1,{\bf r}_2,t)=N^{2}P_{2}'({\bf r}_1,{\bf r}_2,t)$, we get at the order $1/N$:
\begin{eqnarray}
\frac{\partial\omega}{\partial t}({\bf r}_1,t)+\frac{N-1}{N}\langle {\bf V}\rangle({\bf r}_1,t)\cdot \frac{\partial \omega}{\partial {\bf r}_{1}}({\bf r}_1,t)=-\gamma
\frac{\partial}{\partial {\bf r}_{1}}\cdot \int {\bf V}(2\rightarrow 1)g({\bf r}_{1},{\bf r}_{2},t)\, d{\bf r}_{2},
\label{bv7}
\end{eqnarray}
\begin{eqnarray}
\label{bv8} \frac{1}{2}\frac{\partial g}{\partial t}({\bf r}_1,{\bf r}_2,t)+\langle {\bf V}\rangle({\bf r}_1,t)\cdot \frac{\partial g}{\partial {\bf r}_{1}}({\bf r}_1,{\bf r}_2,t)+\left\lbrack\frac{1}{\gamma}\int  {\bf V}(3\rightarrow
1)g({\bf r}_{2},{\bf r}_{3},t) \, d{\bf
r}_{3}\right\rbrack \frac{\partial \omega}{\partial {\bf r}_{1}}({\bf r}_1,t)\nonumber\\
+\frac{1}{\gamma^{2}}
 \tilde{\bf V}(2\rightarrow 1)\cdot \frac{\partial}{\partial {\bf r}_1}\omega({\bf r}_1,t)\omega({\bf r}_2,t)+(1\leftrightarrow 2)=0.
\end{eqnarray}
We have introduced the mean velocity in ${\bf r}_1$ created by all the vortices
\begin{eqnarray}
\label{bv9}
\langle {\bf V}\rangle({\bf r}_1,t) =\frac{1}{\gamma}\int {\bf V}(2\rightarrow 1)\omega({\bf r}_2,t)\, d{\bf r}_{2}=-{\bf z}\times \nabla\psi({\bf r}_1,t),
\end{eqnarray}
and the fluctuating velocity created by point vortex $2$ on point
vortex $1$:
\begin{eqnarray}
\label{bv10}
\tilde{\bf V}(2\rightarrow 1)={\bf V}(2\rightarrow 1)-\frac{1}{N}\langle {\bf V}\rangle({\bf r}_1,t).
\end{eqnarray}
We also recall that the exact velocity created by point vortex $2$ on point
vortex $1$ can be written
\begin{eqnarray}
\label{bv11}
{\bf V}(2\rightarrow 1)=-\gamma\, {\bf z}\times \frac{\partial u_{12}}{\partial {\bf r}_1},
\end{eqnarray}
where $u_{12}=u(|{\bf r}_1-{\bf r}_2|)$ is the binary potential of interaction between point vortices.
Equations (\ref{bv7}) and (\ref{bv8}) are exact at the order $O(1/N)$. They form the right basis to develop a kinetic theory of point vortices at this order of approximation.

\subsection{The limit $N\rightarrow +\infty$: the 2D Euler equation (collisionless regime)}
\label{sec_eu}

In the limit $N\rightarrow +\infty$ for a fixed time $t$, the
correlations between point vortices can be neglected and the $N$-body distribution factorizes in a product of $N$
one-body distributions:
\begin{eqnarray}
\label{eu1}
P_{N}({\bf r}_{1},...,{\bf r}_{N},t)=\prod_{i=1}^{N}P_{1}({\bf r}_{i},t).
\end{eqnarray}
Therefore, for long-range interactions, the mean field approximation is exact at the thermodynamic limit $N\rightarrow +\infty$. Substituting the factorization (\ref{eu1}) in the Liouville equation (\ref{bv1}), and integrating on ${\bf r}_2$, ${\bf r}_3$, ..., ${\bf r}_N$ we find that the smooth vorticity field
$\omega({\bf r},t)$ of the point vortex gas is solution of the 2D
Euler equation
\begin{eqnarray}
\label{eu2}
\frac{\partial \omega}{\partial t}({\bf r}_1,t)+\langle {\bf
V}\rangle({\bf r}_1,t)\cdot \frac{\partial\omega}{\partial {\bf r}_1}({\bf r}_1,t)=0.\label{h4ff}
\end{eqnarray}
This equation also results from Eq. (\ref{bv7}) if we neglect the
correlation function $g({\bf r}_1,{\bf r}_2,t)$ in the right-hand side. The 2D Euler
equation describes the {\it collisionless} evolution of the point
vortex gas for times smaller than $N t_D$. In practice, $N$ is large so that the domain of validity of the 2D Euler equation is huge. The 2D Euler equation is the counterpart of the Vlasov equation in plasma physics and stellar dynamics. It can undergo a process of
mixing and violent relaxation towards a quasistationary state (QSS)
on a very short timescale, of the order of a few dynamical times
$t_{D}$. This QSS has the form of a large-scale vortex.  Miller \cite{miller} and Robert \& Sommeria \cite{rs} have developed a statistical mechanics of the 2D Euler equation to predict these QSSs. The MRS theory is the counterpart of the
Lynden-Bell theory for collisionless stellar systems  \cite{lb}. The analogy between two-dimensional vortices and stellar systems is developed in \cite{houchesPH}.

\subsection{The order $O(1/N)$: an exact kinetic equation (collisional regime)}
\label{sec_dev}

If we want to describe the {\it collisional} evolution of the point vortex gas, we need to consider finite $N$ effects. Equations (\ref{bv7}) and (\ref{bv8}) are valid at the order $1/N$ so they describe the evolution of the system on a timescale of order $Nt_D$. The equation for the evolution of the smooth vorticity field is of the form
\begin{eqnarray}
\frac{\partial\omega}{\partial t}({\bf r}_1,t)+\frac{N-1}{N}\langle {\bf V}\rangle({\bf r}_1,t)\cdot \frac{\partial \omega}{\partial {\bf r}_{1}}({\bf r}_1,t)=C\left\lbrack \omega({\bf r}_1,t)\right\rbrack,
\label{dev0}
\end{eqnarray}
where $C(\omega)$ is a ``collision'' term analogous to the one arising in the Boltzmann equation in the theory of gases. In the present context, there are no real collisions between point vortices. The term on the right-hand side of Eq. (\ref{bv7}) is due to the development of correlations between vortices. It is induced by the two-body correlation function $g({\bf r}_1,{\bf r}_2,t)$ which is determined in terms of the vorticity by Eq. (\ref{bv8}). Our aim is to derive a kinetic equation that is valid at the order $1/N$ and that gives the first correction to the Euler equation.

The formal solution to  Eq. (\ref{bv8}) is
\begin{eqnarray}
\label{dev1} g({\bf r}_1,{\bf r}_2,t)=-\frac{1}{\gamma^2}\int d{\bf r}'_1\int d{\bf r}'_2\int_{0}^{t}dt' U({\bf r}_1,{\bf r}'_1,t-t')U({\bf r}_2,{\bf r}'_2,t-t')\nonumber\\
\times \left\lbrack
 \tilde{\bf V}(2'\rightarrow 1')\cdot \frac{\partial}{\partial {\bf r}'_1}+\tilde{\bf V}(1'\rightarrow 2')\cdot \frac{\partial}{\partial {\bf r}'_2}\right\rbrack
 \omega({\bf r}'_1,t')\omega({\bf r}'_2,t'),
\end{eqnarray}
where the propagator $U({\bf r}_1,{\bf r}'_1,t-t')$ satisfies the equation
\begin{eqnarray}
\label{dev2}
\frac{\partial U}{\partial t}({\bf r}_1,{\bf r}'_1,t-t')+\langle {\bf
V}\rangle({\bf r}_1,t)\cdot \frac{\partial}{\partial {\bf r}_1}U({\bf r}_1,{\bf r}'_1,t-t')+\left\lbrack\int {\bf V}(2\rightarrow 1) U({\bf r}_2,{\bf r}'_1,t-t')\, d{\bf r}_2\right\rbrack\frac{\partial \omega}{\partial {\bf r}_1}({\bf r}_1,t)=0,
\end{eqnarray}
with the initial condition $U({\bf r}_1,{\bf r}'_1,0)=\delta({\bf r}_1-{\bf r}'_1)$. Equation (\ref{dev2}) can be viewed as a linearized version of the 2D Euler equation (see Appendix \ref{sec_euler}). Indeed, if we make the replacement $\omega\rightarrow \omega+\delta\omega$ in Eq. (\ref{euler1}) and linearize it with respect to $\delta\omega$ [see Eq. (\ref{euler2})], we obtain Eq. (\ref{dev2}) in which $U$ plays the role of $\delta\omega$. Consequently, the propagator $U$ obeys the linearized Euler equation.

Substituting Eq. (\ref{dev1}) in Eq. (\ref{bv7}), we obtain a kinetic equation
\begin{eqnarray}
\label{dev3}
{\frac{\partial \omega}{\partial t}}({\bf r}_1,t)+\frac{N-1}{N}\langle {\bf
V}\rangle({\bf r}_1,t)\cdot \frac{\partial\omega}{\partial {\bf r}_1}({\bf r}_1,t)
=\frac{1}{\gamma}\frac{\partial}{\partial {\bf r}_1}\cdot \int d{\bf r}_2\int d{\bf r}'_1\int d{\bf r}'_2\int_{0}^{t}d\tau U({\bf r}_1,{\bf r}'_1,\tau)U({\bf r}_2,{\bf r}'_2,\tau)\nonumber\\
\times  {\bf V}(2\rightarrow 1)\left\lbrack
 \tilde{\bf V}(2'\rightarrow 1')\cdot \frac{\partial}{\partial {\bf r}'_1}+\tilde{\bf V}(1'\rightarrow 2')\cdot \frac{\partial}{\partial {\bf r}'_2}\right\rbrack
 \omega({\bf r}'_1,t-\tau)\omega({\bf r}'_2,t-\tau),
\end{eqnarray}
that is exact at the order $1/N$. If we neglect collective effects, we
recover the generalized Landau equation (\ref{nocoll3}) that was derived in our previous articles
\cite{pre,bbgky,kindetail}. Equation (\ref{dev3}) is a complicated
non-Markovian integrodifferential equation. It is furthermore coupled
to Eq. (\ref{dev2}) which determines the evolution of the
propagator. In order to resolve this coupling, it is necessary to
consider the timescales involved in the dynamics. We shall argue that,
for a given vorticity profile $\omega$, the two-body correlation
function relaxes to its asymptotic form on a timescale short compared
with that on which $\omega$ changes appreciably. This is the
equivalent of the Bogoliubov ansatz in plasma physics. It is expected
to be a very good approximation for $N\gg 1$ since the two-body
correlation function relaxes on a few dynamical times $t_D$ while the
vorticity field changes on a collisional relaxation time of the order
$Nt_D$ or larger. Therefore, it is possible to neglect the time
variation of $\omega({\bf r},t-\tau)$ in the calculation of the
collision term and extend the time integration to $+\infty$. This
amounts to replacing the two-body correlation function in
Eq. (\ref{bv7}) by its asymptotic value $g({\bf r}_1,{\bf
r}_2,+\infty)$ for a given vorticity profile $\omega$. After the correlation function has been obtained as a
functional of $\omega$, the time dependence of $\omega$ can be
reinserted. With this Bogoliubov ansatz (or adiabatic hypothesis), the
kinetic equation (\ref{dev3}) can be rewritten
\begin{eqnarray}
\label{dev4}
{\frac{\partial \omega}{\partial t}}({\bf r}_1,t)+\frac{N-1}{N}\langle {\bf
V}\rangle({\bf r}_1,t)\cdot \frac{\partial\omega}{\partial {\bf r}_1}({\bf r}_1,t)
=\frac{1}{\gamma}\frac{\partial}{\partial {\bf r}_1}\cdot \int d{\bf r}_2\int d{\bf r}'_1\int d{\bf r}'_2\int_{0}^{+\infty}d\tau U({\bf r}_1,{\bf r}'_1,\tau)U({\bf r}_2,{\bf r}'_2,\tau)\nonumber\\
\times {\bf V}(2\rightarrow 1)\left\lbrack
 \tilde{\bf V}(2'\rightarrow 1')\cdot \frac{\partial}{\partial {\bf r}'_1}+\tilde{\bf V}(1'\rightarrow 2')\cdot \frac{\partial}{\partial {\bf r}'_2}\right\rbrack
 \omega({\bf r}'_1,t)\omega({\bf r}'_2,t).
\end{eqnarray}
Similarly, the equation for the propagator takes the form
\begin{eqnarray}
\label{dev5}
\frac{\partial U}{\partial \tau}({\bf r}_1,{\bf r}'_1,\tau)+\langle {\bf
V}\rangle({\bf r}_1,t)\cdot \frac{\partial}{\partial {\bf r}_1}U({\bf r}_1,{\bf r}'_1,\tau)+\left\lbrack\int {\bf V}(2\rightarrow 1) U({\bf r}_2,{\bf r}'_1,\tau)\, d{\bf r}_2\right\rbrack\frac{\partial \omega}{\partial {\bf r}_1}({\bf r}_1,t)=0,
\end{eqnarray}
with the initial condition $U({\bf r}_1,{\bf r}'_1,0)=\delta({\bf r}_1-{\bf r}'_1)$. The two equations (\ref{dev4}) and (\ref{dev5}) are now completely decoupled. For a {\it given} vorticity profile $\omega({\bf r},t)$ at time $t$, one can solve Eq. (\ref{dev5}) to obtain $U({\bf r}_1,{\bf r}'_1,\tau)$ and  determine the collision term in the right-hand side of Eq. (\ref{dev4}). Then, the vorticity profile $\omega({\bf r},t)$ evolves with time on a slow timescale according to Eq. (\ref{dev4}). Interestingly, the structure of this kinetic equation bears a clear physical meaning in terms of generalized Kubo relations \cite{bbgky}. This equation is valid at the order $1/N$ and, for $N\rightarrow +\infty$, it reduces to the (smooth) 2D Euler equation (\ref{eu2})  which describes the collisionless evolution of the point vortex gas.

The kinetic  equation (\ref{dev4}) is, of course, equivalent to the pair of equations
\begin{eqnarray}
\frac{\partial\omega}{\partial t}({\bf r}_1,t)+\frac{N-1}{N}\langle {\bf V}\rangle({\bf r}_1,t)\cdot \frac{\partial \omega}{\partial {\bf r}_{1}}({\bf r}_1,t)=-\gamma
\frac{\partial}{\partial {\bf r}_{1}}\cdot \int {\bf V}(2\rightarrow 1)g({\bf r}_{1},{\bf r}_{2},+\infty)\, d{\bf r}_{2},
\label{dev6}
\end{eqnarray}
\begin{eqnarray}
\label{dev7} g({\bf r}_1,{\bf r}_2,+\infty)=-\frac{1}{\gamma^2}\int d{\bf r}'_1\int d{\bf r}'_2\int_{0}^{+\infty}d\tau U({\bf r}_1,{\bf r}'_1,\tau)U({\bf r}_2,{\bf r}'_2,\tau)\nonumber\\
\times\left\lbrack
 {\bf V}(2'\rightarrow 1')\cdot \frac{\partial}{\partial {\bf r}'_1}+{\bf V}(1'\rightarrow 2')\cdot \frac{\partial}{\partial {\bf r}'_2}\right\rbrack
 \omega({\bf r}'_1,t)\omega({\bf r}'_2,t).
\end{eqnarray}
which correspond to the first two equations of the BBGKY-like hierarchy at the order $1/N$  within the Bogoliubov ansatz. These equations, supplemented by Eq. (\ref{dev5}) for the propagator, provide the formal solution of the problem in the general case. In order to obtain more explicit expressions, we have to consider particular types of flow.

\section{Explicit kinetic equation for axisymmetric flows}
\label{sec_axi}

\subsection{Laplace-Fourier transforms}
\label{sec_lf}

We consider an axisymmetric distribution of point vortices that  is a stable steady state of the 2D Euler equation. Therefore, the vorticity field evolves in time only because of the development of correlations between point vortices due to finite $N$ effects (graininess). In that case, an explicit form of the kinetic equation can be derived.

For an axisymmetric flow, introducing a system of polar coordinates, the vorticity field and the two-body correlation function can be written as $\omega({\bf r}_1,t)=\omega(r_1,t)$ and $g({\bf r}_1,{\bf r}_2,t)=g({r}_1,{r}_2,\theta_1-\theta_2,t)$, and the mean velocity as $\langle {\bf V}\rangle({\bf r}_1,t)=\langle V\rangle_{\theta}(r_1,t){\bf e}_{\theta}$. On the other hand, according to Eq. (\ref{bv11}), the radial velocity (in the direction of ${\bf r}_1$) created by point vortex $2$ on point vortex $1$, is
\begin{eqnarray}
\label{axi1}
V_{r_1}(2\rightarrow 1)=\frac{\gamma}{r_1}\frac{\partial u_{12}}{\partial\theta_1}=-\frac{r_2}{r_1}V_{r_2}(1\rightarrow 2),
\end{eqnarray}
where $u_{12}=u(r_1,r_2,\theta_1-\theta_2)$ is symmetric in $r_1$ and $r_2$ and even in $\phi=\theta_1-\theta_2$ (see Appendix \ref{sec_int}). In that case, Eqs. (\ref{dev6}) and (\ref{dev7}) take the form
\begin{eqnarray}
\frac{\partial\omega}{\partial t}(r_1,t)=-\gamma^2\frac{1}{r_1}\frac{\partial}{\partial r_1}\int_0^{+\infty} r_2\, dr_2\int_{0}^{2\pi} d\theta_2\, \frac{\partial u}{\partial \theta_1}(r_1,r_2,\theta_1-\theta_2) g(r_1,r_2,\theta_1-\theta_2,+\infty),
\label{axi2}
\end{eqnarray}
\begin{eqnarray}
\label{axi3} g({r}_1,{r}_2,\theta_1-\theta_2,+\infty)=-\frac{1}{\gamma}\int r'_1 d{r}'_1 d\theta'_1\int r'_2 d{r}'_2 d\theta'_2\int_{0}^{+\infty}d\tau \frac{\partial u}{\partial\theta_1}(r'_1,r'_2,\theta'_1-\theta'_2)\left\lbrack\left (\frac{1}{r'_1}\frac{\partial}{\partial {r}'_1}-\frac{1}{r'_2}\frac{\partial}{\partial {r}'_2}\right )\omega({r}'_1)\omega({r}'_2)\right\rbrack
  \nonumber\\
 \times U({r}_1,{r}'_1,\theta_1-\theta'_1,\tau)U({r}_2,{r}'_2,\theta_2-\theta'_2,\tau).\quad
\end{eqnarray}
For convenience, we have not written the time $t$ in the vorticity field $\omega(r,t)$ appearing in the correlation function. As we have previously explained, the vorticity profile is assumed ``frozen'' on the short timescale that we consider to compute the asymptotic expression of the correlation function and the collision term (Bogoliubov ansatz). The time $t$ will be restored at the end in the kinetic equation.

We now expand the potential of interaction in Fourier series
\begin{eqnarray}
\label{axi4}
u(r,r',\theta-\theta')=\sum_n e^{in(\theta-\theta')}\hat{u}_n(r,r'),
\end{eqnarray}
and perform similar expansions for $g({r}_1,{r}_2,\theta_1-\theta_2)$ and $U({r}_1,{r}'_1,\theta_1-\theta'_1,t)$. In terms of these Fourier transforms, Eqs. (\ref{axi2}) and (\ref{axi3}) can be rewritten
\begin{eqnarray}
\label{axi5} \frac{\partial\omega}{\partial t}(r_1,t)=2i\pi\gamma^2\frac{1}{r_1}\frac{\partial}{\partial r_1}\int_0^{+\infty}r_2\, dr_2\sum_n  n \, \hat{u}_n(r_1,r_2)\hat{g}_n(r_1,r_2,+\infty),
\end{eqnarray}
\begin{eqnarray}
\label{axi6} \hat{g}_n({r}_1,{r}_2,+\infty)=-i\frac{(2\pi)^2}{\gamma}\int_0^{+\infty} r'_1 d{r}'_1 \int_0^{+\infty} r'_2 d{r}'_2 \int_{0}^{+\infty}d\tau \,  n \, \hat{u}_n(r'_1,r'_2)\left\lbrack\left (\frac{1}{r'_1}\frac{\partial}{\partial {r}'_1}-\frac{1}{r'_2}\frac{\partial}{\partial {r}'_2}\right )\omega({r}'_1)\omega({r}'_2)\right\rbrack
  \nonumber\\
 \times U_n({r}_1,{r}'_1,\tau)U_{-n}({r}_2,{r}'_2,\tau).
\end{eqnarray}
Introducing the Laplace transform of $U_n({r}_1,{r}'_1,\tau)$ (see Appendix \ref{sec_euler} for the definition of Laplace transforms) and integrating on time $\tau$, we get
\begin{eqnarray}
\label{axi7} \hat{g}_n({r}_1,{r}_2,+\infty)=-\frac{1}{\gamma}\int_0^{+\infty} r'_1 d{r}'_1 \int_0^{+\infty} r'_2 d{r}'_2 \int_{\cal C} {d\sigma}\int_{\cal C} {d\sigma'} \frac{1}{\sigma+\sigma'}   n \hat{u}_n(r'_1,r'_2)\left\lbrack\left (\frac{1}{r'_1}\frac{\partial}{\partial {r}'_1}-\frac{1}{r'_2}\frac{\partial}{\partial {r}'_2}\right )\omega({r}'_1)\omega({r}'_2)\right\rbrack
  \nonumber\\
 \times U_n({r}_1,{r}'_1,\sigma)U_{-n}({r}_2,{r}'_2,\sigma'),
\end{eqnarray}
where ${\cal C}$ is the Laplace contour in the complex $\sigma$-plane. The integration over $\sigma'$ can be performed by closing the contour by an infinite semicircle in the upper half-plane. Since $U_{-n}({r}_2,{r}'_2,\sigma')$ vanishes for $|\sigma'|\rightarrow +\infty$, the only contribution of the integral comes from the pole at $\sigma'=-\sigma$. Using the residue theorem, we obtain
\begin{eqnarray}
\label{axi8} \hat{g}_n({r}_1,{r}_2,+\infty)=-\frac{2\pi i}{\gamma}\int_0^{+\infty} r'_1 d{r}'_1 \int_0^{+\infty} r'_2 d{r}'_2 \int_{-\infty}^{+\infty} d\sigma \, n \hat{u}_n(r'_1,r'_2)\left\lbrack\left (\frac{1}{r'_1}\frac{\partial}{\partial {r}'_1}-\frac{1}{r'_2}\frac{\partial}{\partial {r}'_2}\right )\omega({r}'_1)\omega({r}'_2)\right\rbrack
  \nonumber\\
 \times U_n({r}_1,{r}'_1,\sigma)U_{-n}({r}_2,{r}'_2,-\sigma).
\end{eqnarray}
Finally, substituting Eq. (\ref{axi8}) in Eq. (\ref{axi5}), the kinetic equation takes the form
\begin{eqnarray}
\label{axi9} \frac{\partial\omega}{\partial t}(r_1,t)=(2\pi)^2\gamma\frac{1}{r_1}\frac{\partial}{\partial r_1}\int_0^{+\infty}r_2\, dr_2\sum_n  n \hat{u}_n(r_1,r_2)\int_0^{+\infty} r'_1 d{r}'_1 \int_0^{+\infty} r'_2 d{r}'_2 \int_{-\infty}^{+\infty} d\sigma \, n \hat{u}_n(r'_1,r'_2)\nonumber\\
\times\left\lbrack\left (\frac{1}{r'_1}\frac{\partial}{\partial {r}'_1}-\frac{1}{r'_2}\frac{\partial}{\partial {r}'_2}\right )\omega({r}'_1)\omega({r}'_2)\right\rbrack U_n({r}_1,{r}'_1,\sigma)U_{-n}({r}_2,{r}'_2,-\sigma).
\end{eqnarray}
On the other hand, for an axisymmetric vorticity distribution $\omega(r)$, the Laplace-Fourier transform of the propagator is explicitly given by (see Appendix \ref{sec_euler}):
\begin{eqnarray}
U_n(r,r',\sigma)=\frac{i}{\sigma-n\Omega(r)}\frac{\delta(r-r')}{2\pi r}+i\frac{G(n,r,r',\sigma)}{(\sigma-n\Omega(r))(\sigma-n\Omega(r'))}n\frac{1}{r}\frac{\partial\omega}{\partial r}(r).
 \label{axi10}
\end{eqnarray}
The first term on the right-hand side represents the advection by the mean flow, i.e. it corresponds to a pure rotation with angular velocity $\Omega(r)$. The second term takes into account collective effects.

Before going further, some technical details must be given.  Since $U_n({r}_1,{r}'_1,\sigma)$ is obtained as the Laplace transform of the propagator $U_n({r}_1,{r}'_1,\tau)$, this function is analytic in the upper half of the complex $\sigma$ plane. It is then continued analytically into the lower half-plane where it generally has singularities. The contour of $\sigma$ integration in Eq. (\ref{axi9}) sees all singularities of $U_n({r}_1,{r}'_1,\sigma)$ from above. On the other hand, the function $U_{-n}({r}_1,{r}'_1,-\sigma)$ which is the complex conjugate to $U_n({r}_1,{r}'_1,\sigma)$ is an analytic function in the lower half-plane and is continued analytically into the upper half-plane where it generally has singularities. The integration contour in Eq. (\ref{axi9}) sees all the singularities of  $U_{-n}({r}_1,{r}'_1,-\sigma)$ from below. In order to take these boundary conditions into account, we shall write $\sigma+i0^+$ in place of $\sigma$ for the functions which are well defined and analytic in the upper half-plane and we use $\sigma-i0^+$ for the functions well defined and analytic in the lower half-plane.

\subsection{Without collective effects: Landau-type equation}
\label{sec_lan}

Before solving the general case, we first derive a Landau-type equation obtained by neglecting collective effects. In that case, the propagator (\ref{axi10}) reduces to
\begin{eqnarray}
U_n(r,r',\sigma)=\frac{i}{\sigma-n\Omega(r)}\frac{\delta(r-r')}{2\pi r},
 \label{lan1}
\end{eqnarray}
and the kinetic equation (\ref{axi9}) becomes
\begin{eqnarray}
\label{lan2} \frac{\partial\omega}{\partial t}(r_1,t)=\gamma\frac{1}{r_1}\frac{\partial}{\partial r_1}\int_0^{+\infty}r_2\, dr_2\int_{-\infty}^{+\infty} d\sigma \, \sum_n  n^2 \hat{u}_n(r_1,r_2)^2 \frac{1}{(\sigma-n\Omega(r_1)+i0^+)(\sigma-n\Omega(r_2)-i0^+)} \nonumber\\
\times \left (\frac{1}{r_1}\frac{\partial}{\partial {r}_1}-\frac{1}{r_2}\frac{\partial}{\partial {r}_2}\right )\omega({r}_1)\omega({r}_2).
\end{eqnarray}
The integration on $\sigma$ may be carried out by closing the contour with an infinite semicircle in the lower half-plane. Only the pole at $\sigma=n\Omega(r_1)$ contributes. Using the residue theorem, we obtain
\begin{eqnarray}
\label{lan3} \frac{\partial\omega}{\partial t}(r_1,t)=-2\pi i \gamma\frac{1}{r_1}\frac{\partial}{\partial r_1}\int_0^{+\infty}r_2\, dr_2 \, \sum_n  n^2 \hat{u}_n(r_1,r_2)^2 \frac{1}{n\Omega(r_1)-n\Omega(r_2)-i0^+} \left (\frac{1}{r_1}\frac{\partial}{\partial {r}_1}-\frac{1}{r_2}\frac{\partial}{\partial {r}_2}\right )\omega({r}_1)\omega({r}_2).
\end{eqnarray}
Then, with the aid of the Plemelj formula
\begin{eqnarray}
\label{lan4} \frac{1}{x\pm i0^+}={\cal P}\left (\frac{1}{x}\right )\mp i\pi \delta(x),
\end{eqnarray}
the foregoing expression can be rewritten
\begin{eqnarray}
\label{lan5} \frac{\partial\omega}{\partial t}(r_1,t)=2\pi^2 \gamma\frac{1}{r_1}\frac{\partial}{\partial r_1}\int_0^{+\infty}r_2\, dr_2 \, \sum_n  n^2 \hat{u}_n(r_1,r_2)^2 \delta\lbrack n(\Omega(r_1)-\Omega(r_2))\rbrack\left (\frac{1}{r_1}\frac{\partial}{\partial {r}_1}-\frac{1}{r_2}\frac{\partial}{\partial {r}_2}\right )\omega({r}_1)\omega({r}_2).
\end{eqnarray}
Finally, using the identity $\delta(\lambda x)=\frac{1}{|\lambda|}\delta(x)$ and putting back the (slow) time dependence in the kinetic equation, we get
\begin{eqnarray}
\label{lan6} \frac{\partial\omega}{\partial t}(r_1,t)=2\pi^2 \gamma\frac{1}{r_1}\frac{\partial}{\partial r_1}\int_0^{+\infty}r_2\, dr_2 \, \sum_n  |n| \hat{u}_n(r_1,r_2)^2 \delta \lbrack \Omega(r_1,t)-\Omega(r_2,t)\rbrack\left (\frac{1}{r_1}\frac{\partial}{\partial {r}_1}-\frac{1}{r_2}\frac{\partial}{\partial {r}_2}\right )\omega({r}_1,t)\omega({r}_2,t).
\end{eqnarray}
Using Eq. (\ref{int3a}), the series can be explicitly calculated and we obtain the alternative form
\begin{eqnarray}
\frac{\partial \omega}{\partial t}(r_1,t)=2\pi^{2}{\gamma} \frac{1}{r_1}
\frac{\partial}{\partial r_1} \int_0^{+\infty} r_2 dr_2\,
\chi(r_{1},r_{2})\delta\left\lbrack \Omega(r_1,t)-\Omega(r_2,t)\right\rbrack \left
(\frac{1}{r_1}\frac{\partial}{\partial
r_1}-\frac{1}{r_2}\frac{\partial}{\partial
 r_2}\right )\omega({r}_1,t)\omega({r}_2,t),\label{lan7}
\end{eqnarray}
with
\begin{eqnarray}
\chi(r_{1},r_{2})=\sum_n  |n| \hat{u}_n(r_1,r_2)^2=\frac{1}{8\pi^{2}}\sum_{m=1}^{+\infty}\frac{1}{m}\left (\frac{r_{<}}{r_{>}}\right )^{2m}=-\frac{1}{8\pi^{2}}
\ln\left \lbrack 1-\left (\frac{r_{<}}{r_{>}}\right )^{2}\right\rbrack,
\label{lan8}
\end{eqnarray}
where $r_{>}={\rm max}(r_{1},r_{2})$ and $r_{<}={\rm
min}(r_{1},r_{2})$. This kinetic equation, which neglects collective
effects and takes into account only two-body encounters, is the
counterpart of the Landau \cite{landau} equation in plasma physics. It
was derived in \cite{pre,bbgky,kindetail} from different
formalisms. In plasma physics, the Landau equation presents a
logarithmic divergence at large scales. Therefore, collective effects
which lead to Debye shielding, are crucial because they regularize the
logarithmic divergence at large scales in the Landau equation. This is
essentially what the works of Lenard \cite{lenard} and Balescu
\cite{balescu} have demonstrated. Since the kinetic equation
(\ref{lan6}) does not present any divergence, the neglect of
collective effects may not be crucial for point vortices.

\subsection{With collective effects: Lenard-Balescu-type equation}
\label{sec_lb}

If we take collective effects into account, the kinetic equation is
obtained by substituting the expression (\ref{axi10}) of the
propagator in Eq. (\ref{axi9}) and carrying out the integrations. The calculations
are similar to those developed by Ichimaru \cite{ichimaru} in his derivation of the
Lenard-Balescu equation from the BBGKY hierarchy.

Let us first perform the integration on $r_2$. Using Eq. (\ref{axi10}), we obtain
\begin{eqnarray}
2\pi \int_0^{+\infty} r_2 dr_2\, \hat{u}_n(r_1,r_2) U_{-n}(r_2,r'_2,-\sigma)= \int_0^{+\infty} r_2 dr_2\, \hat{u}_n(r_1,r_2) \nonumber\\
\times\left\lbrace\frac{i}{-\sigma+n\Omega(r_2)}\frac{\delta(r_2-r'_2)}{r_2}-2\pi i
\frac{ G(-n,r_2,r'_2,-\sigma)}{(-\sigma+n\Omega(r_2))(-\sigma+n\Omega(r'_2))}n\frac{1}{r_2}
\frac{\partial\omega}{\partial r_2}(r_2)\right\rbrace,
 \label{lb1}
\end{eqnarray}
which can be rewritten
\begin{eqnarray}
2\pi \int_0^{+\infty} r_2 dr_2\, \hat{u}_n(r_1,r_2) U_{-n}(r_2,r'_2,-\sigma)=\frac{i}{-\sigma+n\Omega(r'_2)}\nonumber\\
\times\left\lbrack \hat{u}_n(r_1,r'_2)-2\pi  \int_0^{+\infty} dr_2\, \hat{u}_n(r_1,r_2)
\frac{ G(-n,r_2,r'_2,-\sigma)}{-\sigma+n\Omega(r_2)}n
\frac{\partial\omega}{\partial r_2}(r_2)\right\rbrack.
 \label{lb2}
\end{eqnarray}
Using the identity (\ref{int6}), this integral is finally reduced to the form
\begin{eqnarray}
2\pi \int_0^{+\infty} r_2 dr_2\, \hat{u}_n(r_1,r_2) U_{-n}(r_2,r'_2,-\sigma)=\frac{i}{-\sigma+n\Omega(r'_2)}
 G(-n,r_1,r'_2,-\sigma),
 \label{lb3}
\end{eqnarray}
where $G(n,r_1,r_2,\sigma)$ is the ``dressed'' potential of interaction between point vortex $1$ and point vortex $2$. It is solution of the differential equation (\ref{euler13}).

Substituting Eqs. (\ref{axi10}) and (\ref{lb3}) in the kinetic equation (\ref{axi9}), we obtain
\begin{eqnarray}
\label{lb4} \frac{\partial\omega}{\partial t}(r_1,t)=2\pi\gamma\frac{1}{r_1}\frac{\partial}{\partial r_1}\sum_n  n \int_0^{+\infty} r'_1 d{r}'_1 \int_0^{+\infty} r'_2 d{r}'_2 \int_{-\infty}^{+\infty} d\sigma \, n \hat{u}_n(r'_1,r'_2)\left\lbrack\left (\frac{1}{r'_1}\frac{\partial}{\partial {r}'_1}-\frac{1}{r'_2}\frac{\partial}{\partial {r}'_2}\right )\omega({r}'_1)\omega({r}'_2)\right\rbrack\nonumber\\
\left\lbrack \frac{1}{\sigma-n\Omega(r_1)+i0^+}\frac{\delta(r_1-r'_1)}{2\pi r_1}+\frac{ G(n,r_1,r'_1,\sigma)}{(\sigma-n\Omega(r_1)+i0^+)(\sigma-n\Omega(r'_1)+i0^+)}n\frac{1}{r_1}
\frac{\partial\omega}{\partial r_1}(r_1)\right\rbrack\nonumber\\
\times\frac{1}{\sigma-n\Omega(r'_2)-i0^+}
 G(-n,r_1,r'_2,-\sigma).
\end{eqnarray}
This equation can be decomposed as follows
\begin{eqnarray}
\label{lb5} \frac{\partial\omega}{\partial t}(r_1,t)=\gamma\frac{1}{r_1}\frac{\partial}{\partial r_1}\sum_n  n  \int_0^{+\infty} r'_2 d{r}'_2 \int_{-\infty}^{+\infty} d\sigma \, n \hat{u}_n(r_1,r'_2)\left\lbrack\left (\frac{1}{r_1}\frac{\partial}{\partial {r}_1}-\frac{1}{r'_2}\frac{\partial}{\partial {r}'_2}\right )\omega({r}_1)\omega({r}'_2)\right\rbrack\nonumber\\
\times\frac{ G(-n,r_1,r'_2,-\sigma)}{(\sigma-n\Omega(r_1)+i0^+)(\sigma-n\Omega(r'_2)-i0^+)}\nonumber\\
+\gamma\frac{1}{r_1}\frac{\partial}{\partial r_1}\sum_n  n \int_0^{+\infty} r'_2 d{r}'_2 \int_{-\infty}^{+\infty} d\sigma \, n \frac{G(-n,r_1,r'_2,-\sigma)}{(\sigma-n\Omega(r_1)+i0^+)(\sigma-n\Omega(r'_2)-i0^+)}\left\lbrack \frac{1}{r_1}\frac{\partial}{\partial {r}_1}\omega({r}_1)\omega({r}'_2)\right\rbrack\nonumber\\
\times 2\pi\int_0^{+\infty}dr'_1 \hat{u}_n(r'_1,r'_2) n \frac{\partial\omega}{\partial r'_1}(r'_1)\frac{G(n,r_1,r'_1,\sigma)}{\sigma-n\Omega(r'_1)+i0^+}
\nonumber\\
-\gamma\frac{1}{r_1}\frac{\partial}{\partial r_1}\sum_n  n \int_0^{+\infty} r'_1 d{r}'_1 \int_{-\infty}^{+\infty} d\sigma \, n \frac{G(n,r_1,r'_1,\sigma)}{(\sigma-n\Omega(r_1)+i0^+)(\sigma-n\Omega(r'_1)+i0^+)}\left\lbrack \frac{1}{r_1}\frac{\partial}{\partial {r}_1}\omega({r}_1)\omega({r}'_1)\right\rbrack\nonumber\\
\times 2\pi\int_0^{+\infty}dr'_2 \hat{u}_n(r'_1,r'_2) n \frac{\partial\omega}{\partial r'_2}(r'_2)\frac{G(-n,r_1,r'_2,-\sigma)}{\sigma-n\Omega(r'_2)-i0^+}.
\end{eqnarray}
Using the identity (\ref{int6}), we can write more succinctly
\begin{eqnarray}
\label{lb6} \frac{\partial\omega}{\partial t}(r_1,t)=\gamma\frac{1}{r_1}\frac{\partial}{\partial r_1}\sum_n  n  \int_0^{+\infty} r'_2 d{r}'_2 \int_{-\infty}^{+\infty} d\sigma \, n \hat{u}_n(r_1,r'_2)\left\lbrack\left (\frac{1}{r_1}\frac{\partial}{\partial {r}_1}-\frac{1}{r'_2}\frac{\partial}{\partial {r}'_2}\right )\omega({r}_1)\omega({r}'_2)\right\rbrack\nonumber\\
\times\frac{ G(-n,r_1,r'_2,-\sigma)}{(\sigma-n\Omega(r_1)+i0^+)(\sigma-n\Omega(r'_2)-i0^+)}\nonumber\\
+\gamma\frac{1}{r_1}\frac{\partial}{\partial r_1}\sum_n  n \int_0^{+\infty} r'_2 d{r}'_2 \int_{-\infty}^{+\infty} d\sigma \, n \frac{G(-n,r_1,r'_2,-\sigma)}{(\sigma-n\Omega(r_1)+i0^+)(\sigma-n\Omega(r'_2)-i0^+)}\left\lbrack \frac{1}{r_1}\frac{\partial}{\partial {r}_1}\omega({r}_1)\omega({r}'_2)\right\rbrack\nonumber\\
\times \left\lbrack -\hat{u}_n(r_1,r'_2)+G(n,r_1,r'_2,\sigma)\right\rbrack
\nonumber\\
-\gamma\frac{1}{r_1}\frac{\partial}{\partial r_1}\sum_n  n \int_0^{+\infty} r'_1 d{r}'_1 \int_{-\infty}^{+\infty} d\sigma \, n \frac{G(n,r_1,r'_1,\sigma)}{(\sigma-n\Omega(r_1)+i0^+)(\sigma-n\Omega(r'_1)+i0^+)}\left\lbrack \frac{1}{r_1}\frac{\partial}{\partial {r}_1}\omega({r}_1)\omega({r}'_1)\right\rbrack\nonumber\\
\times \left\lbrack -\hat{u}_n(r'_1,r_1)+G(-n,r_1,r'_1,-\sigma)\right\rbrack.
\end{eqnarray}
Parts of the first two terms cancel each other. On the other hand, the third term can be split into two parts. Hence, we get
\begin{eqnarray}
\label{lb7} \frac{\partial\omega}{\partial t}(r_1,t)=-\frac{\gamma}{2\pi}\frac{1}{r_1}\frac{\partial}{\partial r_1}\sum_n  n   \int_{-\infty}^{+\infty} d\sigma \, \omega(r_1)\frac{1}{\sigma-n\Omega(r_1)+i0^+}\nonumber\\
\times 2\pi\int_0^{+\infty} d{r}'_2 \hat{u}_n(r_1,r'_2)n\frac{\partial\omega}{\partial {r}'_2}({r}'_2)\frac{ G(-n,r_1,r'_2,-\sigma)}{\sigma-n\Omega(r'_2)-i0^+}\nonumber\\
+\gamma\frac{1}{r_1}\frac{\partial}{\partial r_1}\sum_n  n \int_0^{+\infty} r'_2 d{r}'_2 \int_{-\infty}^{+\infty} d\sigma \, n \frac{G(-n,r_1,r'_2,-\sigma)G(n,r_1,r'_2,\sigma)}{(\sigma-n\Omega(r_1)+i0^+)(\sigma-n\Omega(r'_2)-i0^+)}\left\lbrack \frac{1}{r_1}\frac{\partial}{\partial {r}_1}\omega({r}_1)\omega({r}'_2)\right\rbrack\nonumber\\
+\gamma\frac{1}{r_1}\frac{\partial}{\partial r_1}\sum_n  n \int_0^{+\infty} r'_1 d{r}'_1 \int_{-\infty}^{+\infty} d\sigma \, n \frac{G(n,r_1,r'_1,\sigma)\hat{u}_n(r'_1,r_1)}{(\sigma-n\Omega(r_1)+i0^+)(\sigma-n\Omega(r'_1)+i0^+)}
\left\lbrack \frac{1}{r_1}\frac{\partial}{\partial {r}_1}\omega({r}_1)\omega({r}'_1)\right\rbrack\nonumber\\
-\gamma\frac{1}{r_1}\frac{\partial}{\partial r_1}\sum_n  n \int_0^{+\infty} r'_1 d{r}'_1 \int_{-\infty}^{+\infty} d\sigma \, n \frac{G(n,r_1,r'_1,\sigma)G(-n,r_1,r'_1,-\sigma)}{(\sigma-n\Omega(r_1)+i0^+)(\sigma-n\Omega(r'_1)+i0^+)}\left\lbrack \frac{1}{r_1}\frac{\partial}{\partial {r}_1}\omega({r}_1)\omega({r}'_1)\right\rbrack.
\end{eqnarray}
Using the identity (\ref{int6}), this can be rewritten
\begin{eqnarray}
\label{lb8} \frac{\partial\omega}{\partial t}(r_1,t)=-\frac{\gamma}{2\pi}\frac{1}{r_1}\frac{\partial}{\partial r_1}\sum_n  n   \int_{-\infty}^{+\infty} d\sigma \, \omega(r_1)\frac{1}{\sigma-n\Omega(r_1)+i0^+}\nonumber\\
\times \left\lbrack -\hat{u}_n(r_1,r_1)+G(-n,r_1,r_1,-\sigma)\right\rbrack\nonumber\\
+\gamma\frac{1}{r_1}\frac{\partial}{\partial r_1}\sum_n  n \int_0^{+\infty} r'_2 d{r}'_2 \int_{-\infty}^{+\infty} d\sigma \, n \frac{G(-n,r_1,r'_2,-\sigma)G(n,r_1,r'_2,\sigma)}{(\sigma-n\Omega(r_1)+i0^+)(\sigma-n\Omega(r'_2)-i0^+)}\left\lbrack \frac{1}{r_1}\frac{\partial}{\partial {r}_1}\omega({r}_1)\omega({r}'_2)\right\rbrack\nonumber\\
+\gamma\frac{1}{r_1}\frac{\partial}{\partial r_1}\sum_n  n \int_0^{+\infty} r'_1 d{r}'_1 \int_{-\infty}^{+\infty} d\sigma \, n \frac{G(n,r_1,r'_1,\sigma)\hat{u}_n(r'_1,r_1)}{(\sigma-n\Omega(r_1)+i0^+)(\sigma-n\Omega(r'_1)+i0^+)}
\left\lbrack \frac{1}{r_1}\frac{\partial}{\partial {r}_1}\omega({r}_1)\omega({r}'_1)\right\rbrack\nonumber\\
-\gamma\frac{1}{r_1}\frac{\partial}{\partial r_1}\sum_n  n \int_0^{+\infty} r'_1 d{r}'_1 \int_{-\infty}^{+\infty} d\sigma \, n \frac{G(n,r_1,r'_1,\sigma)G(-n,r_1,r'_1,-\sigma)}{(\sigma-n\Omega(r_1)+i0^+)(\sigma-n\Omega(r'_1)+i0^+)}\left\lbrack \frac{1}{r_1}\frac{\partial}{\partial {r}_1}\omega({r}_1)\omega({r}'_1)\right\rbrack.
\end{eqnarray}
The integration in the first term may be carried out by closing the contour with an infinite semicircle in the lower half-plane; only the pole at $\sigma=n\Omega(r_1)-i0^+$ contributes. Similarly, by closing the contour in the upper half-plane, we find that the third term vanishes. The second and fourth terms may be combined with the aid of the Plemelj formula (\ref{lan4}). We are thus left with
\begin{eqnarray}
\label{lb9} \frac{\partial\omega}{\partial t}(r_1,t)=i\gamma\frac{1}{r_1}\frac{\partial}{\partial r_1}\sum_n  n \omega(r_1)\left\lbrack -\hat{u}_n(r_1,r_1)+G(-n,r_1,r_1,-n\Omega(r_1))\right\rbrack\nonumber\\
+2\pi i \gamma\frac{1}{r_1}\frac{\partial}{\partial r_1}\sum_n  n^2 \int_0^{+\infty} r_2 d{r}_2 \frac{G(-n,r_1,r_2,-n\Omega(r_2))G(n,r_1,r_2,n\Omega(r_2))}{n(\Omega(r_2)-\Omega(r_1))+i 0^+} \frac{1}{r_1}\frac{\partial}{\partial {r}_1}\omega({r}_1)\omega({r}_2).
\end{eqnarray}
Checking the symmetry of the terms with respect to the inversion of $n$, and again using the Plemelj formula (\ref{lan4}), we get
\begin{eqnarray}
\label{lb10} \frac{\partial\omega}{\partial t}(r_1,t)=\gamma\frac{1}{r_1}\frac{\partial}{\partial r_1}\sum_n  n \omega(r_1) {\rm Im} \lbrack G(n,r_1,r_1,n\Omega(r_1))\rbrack\nonumber\\
+2\pi^2 \gamma\frac{1}{r_1}\frac{\partial}{\partial r_1}\sum_n  n^2 \int_0^{+\infty} r_2 d{r}_2 |G(n,r_1,r_2,n\Omega(r_2))|^2 \delta\lbrack n(\Omega(r_2)-\Omega(r_1))\rbrack \frac{1}{r_1}\frac{\partial}{\partial {r}_1}\omega({r}_1)\omega({r}_2),
\end{eqnarray}
where we have used $G(-n,r,r',-\sigma)=G^*(n,r,r',\sigma)$. To determine the first term in this equation, we multiply Eq. (\ref{euler13}) by $2\pi r\, G(-n,r,r',-\sigma)$, integrate over $r$ from $0$ to $+\infty$ and take the imaginary part of the resulting expression using the Plemelj formula (\ref{lan4}). This yields
\begin{equation}
{\rm Im}\,  G(n,r,r,\sigma)=-2\pi^2\int_{0}^{+\infty} dr'  \, n |G(n,r,r',\sigma)|^2 \delta(\sigma-n\Omega(r'))\frac{\partial\omega}{\partial r'}(r').
\label{lb11}
\end{equation}
Substituting this identity in Eq. (\ref{lb10}) and reinserting  the (slow) time dependence in the the kinetic equation, we finally obtain
\begin{eqnarray}
\label{lb12} \frac{\partial\omega}{\partial t}(r_1,t)=2\pi^2 \gamma\frac{1}{r_1}\frac{\partial}{\partial r_1}\int_0^{+\infty}r_2\, dr_2 \, \sum_n  |n| |G(n,r_1,r_2,n\Omega(r_1,t))|^2 \delta \lbrack \Omega(r_1,t)-\Omega(r_2,t)\rbrack \nonumber\\
\times \left (\frac{1}{r_1}\frac{\partial}{\partial {r}_1}-\frac{1}{r_2}\frac{\partial}{\partial {r}_2}\right )\omega({r}_1,t)\omega({r}_2,t).
\end{eqnarray}
This can be rewritten
\begin{eqnarray}
\frac{\partial \omega}{\partial t}(r_1,t)=2\pi^{2}{\gamma} \frac{1}{r_1}
\frac{\partial}{\partial r_1} \int_0^{+\infty} r_2 dr_2\,
\chi(r_{1},r_{2},\Omega(r_1,t))\delta\left\lbrack \Omega(r_1,t)-\Omega(r_2,t)\right\rbrack \left
(\frac{1}{r_1}\frac{\partial}{\partial
r_1}-\frac{1}{r_2}\frac{\partial}{\partial
 r_2}\right )\omega({r}_1,t)\omega({r}_2,t),\label{lb13}
\end{eqnarray}
with
\begin{eqnarray}
\chi(r_{1},r_{2})=\sum_n  |n| |G(n,r_1,r_2,n\Omega(r_1,t))|^2.
\label{lb14}
\end{eqnarray}
This kinetic equation, which properly takes collective effects into account, is the counterpart of the Lenard-Balescu equation in plasma physics. It was derived in \cite{dubin,klim} from the Klimontovich formalism. We have here provided an alternative derivation of this equation from the BBGKY hierarchy.
Note that the Lenard-Balescu-type equation (\ref{lb12})  differs from the Landau-type equation (\ref{lan6}) only by the replacement of the ``bare'' potential of interaction  $\hat{u}_n(r_1,r_2)$ by the ``dressed'' potential of interaction $G(n,r_1,r_2,n\Omega(r_1,t))$ taking into account the contribution of the polarization cloud. This is similar to the case of plasma physics.

{\it Remark:} the Landau-type kinetic equation (\ref{lan6}) and the
Lenard-Balescu-type kinetic equation (\ref{lb12}) are restricted to
axisymmetric flows. By neglecting collective effects, we have obtained
in \cite{pre,bbgky,kindetail} a generalized Landau equation
(\ref{nocoll3}) that is valid for arbitrary flows. Sano \cite{sano}
confirmed our results and attempted to derive a generalized Lenard-Balescu
equation for arbitrary flows. Unfortunately, the final results are
complicated and not fully explicit. Nevertheless, going beyond the
assumption of axisymmetry as in \cite{pre,bbgky,kindetail} and
\cite{sano} is certainly valuable in order to treat more realistic
flows. The general kinetic equation (\ref{dev3}) could be a good
starting point in this direction.

\subsection{Numerical resolution of the kinetic equation: No relaxation towards the Boltzmann distribution}
\label{sec_num}

\begin{figure*}
\centering
  \includegraphics[width=0.5\textwidth]{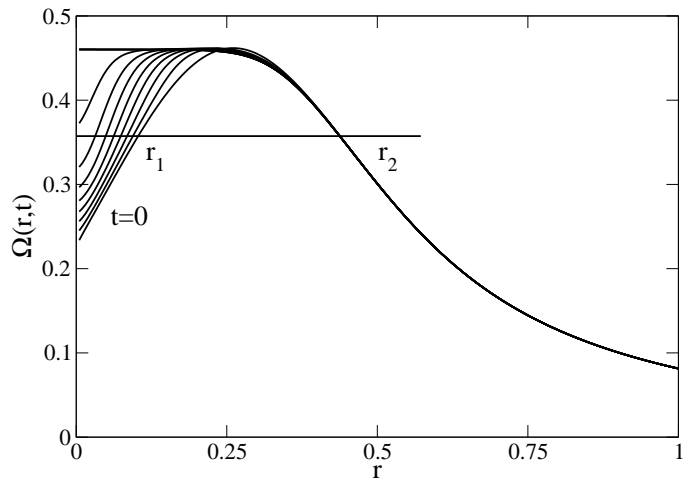}
\caption{Evolution of the profile of angular velocity obtained by solving numerically the kinetic equation (\ref{lan7}) \cite{cl}. Time increases from bottom to top. The horizontal line shows the resonance between point vortices in $r_1$ and $r_2$ which rotate with the same angular velocity. The evolution stops when the profile of angular velocity becomes monotonic so that there is no resonance anymore. The final profile does {\it not} correspond to the Boltzmann distribution  (\ref{nv5}). The kinetic  equation  (\ref{lan7}) is valid on a timescale $Nt_D$. Therefore, on this timescale the vortex gas does not reach statistical equilibrium but remains ``blocked'' in a QSS with a monotonic profile of angular velocity. The Boltzmann distribution  (\ref{nv5}) may be reached on a longer timescale (due to higher order correlations) but this timescale of relaxation is not known. It is not even clear whether the point vortex gas will ever relax towards the statistical equilibrium distribution (\ref{nv5}) in the axisymmetric case. }
\label{fig1}       % Give a unique label
\end{figure*}

The kinetic equation (\ref{lb13}) is valid at the order $1/N$ so it
describes the ``collisional'' evolution of the point vortex gas on a
timescale of order $Nt_D$. This kinetic equation conserves the
circulation $\Gamma$, the energy $E$ and the angular momentum $L$. It
also monotonically increases the Boltzmann entropy in the sense that
$\dot S_{B}\ge 0$ ($H$-theorem). These properties are proven in
\cite{dubin,cl}.  The change of the vorticity distribution in $r_1$ is
due to a condition of resonance (encapsulated in the
$\delta$-function) between vortices located in $r_{1}$ and vortices
located in $r_{2}\neq r_{1}$ which rotate with the {\it same} angular
velocity $\Omega(r_2,t)=\Omega(r_{1},t)$ (the self-interaction at
$r_2=r_1$ does not produce transport since the term in parenthesis
vanishes identically). Of course, this condition can be satisfied only
when the profile of angular velocity is non-monotonic. The collisional
evolution of the point vortices is thus truly due to long-range
interactions since the current in $r_{1}$ is caused by ``distant
collisions'' with vortices located in $r_2\neq r_{1}$ that can be far
away. This is different from the case of plasma physics and stellar
dynamics where the collisions are assumed to be {\it local} in
space. The mean field Boltzmann distribution (\ref{nv5}) is a
particular steady state of Eq. (\ref{lb13}) but it is not the only
one: The kinetic equation (\ref{lb13}) admits an infinite number of
steady states. Indeed, all the vorticity profiles $\omega(r)$
associated with a monotonic profile of angular velocity $\Omega(r)$
are steady states of the kinetic equation (\ref{lb13}) since the
$\delta$-function is zero for these profiles. Therefore, the
collisional evolution of the point vortex gas described by
Eq. (\ref{lb13}) stops when the profile of angular velocity becomes
monotonic (so that there is no resonance) even if the system has not
reached the Boltzmann distribution.  In that case, the system settles
on a QSS that is {\it not} the most mixed state predicted by
statistical mechanics (see Fig. \ref{fig1}) \cite{cl}.  On the
timescale $Nt_D$ on which the kinetic theory is valid, the collisions
tend to create a monotonic profile of angular velocity. Since the
entropy increases monotonically, the vorticity profile {\it tends} to
approach the Boltzmann distribution (the system becomes ``more
mixed'') but does not attain it in general because of the absence of
resonance. This is particularly obvious if we start from an initial
condition with a monotonic profile of angular velocity that is
non-Boltzmannian. In that case, the collision term vanishes, so that
$\partial\omega/\partial t=0$ meaning that there is no evolution on
the timescale $Nt_D$. The Boltzmann distribution may be reached on
longer timescales, larger than $Nt_D$. To describe this regime, we
need to determine terms of order $N^{-2}$, or smaller, in the
expansion of the solutions of the BBGKY hierarchy for $N\rightarrow
+\infty$. This implies in particular the determination of the
three-body, or higher, correlation function, which is a formidable
task. At the moment, we can only conclude from the kinetic theory
that, for an axisymmetric distribution of point vortices, the
relaxation time satisfies
\begin{eqnarray}
\label{relv}
t_{R}>Nt_{D}.
\end{eqnarray}
The collisional relaxation towards the Boltzmann distribution (\ref{nv5}) is
therefore a very slow process. In fact, up to now, there is no rigorous proof coming from the kinetic
theory that the point vortex gas will ever relax towards the Boltzmann
distribution predicted by statistical mechanics. Indeed, the point vortices may not mix well enough through the action of  ``collisions''. Therefore, the relaxation (or not) of the vorticity profile
towards the Boltzmann distribution (\ref{nv5}) for $t\rightarrow +\infty$ still remains an open
problem. This is at variance with the Landau \cite{landau} and
Lenard-Balescu
\cite{lenard,balescu} equations of plasma physics  which always
converge towards the Boltzmann distribution (it is the unique steady
state of these equations).  Therefore, in plasma physics, it is
sufficient to develop the kinetic theory at the order $1/N$. The
kinetic theory of point vortices is more complicated since it requires
to go to higher order in the expansion in power of $1/N$. This is
similar to the case of spatially homogeneous one-dimensional systems
with long-range interactions, such as one-dimensional plasmas
\cite{feix,kp} and the HMF model \cite{bd,cvb}, for which the
Lenard-Balescu collision term also vanishes at the order
$1/N$. Therefore, $t_R>Nt_D$ for these systems. For spatially
homogeneous one-dimensional plasmas, the relaxation time is found
numerically to scale like $t_R\sim N^2 t_D$ \cite{dawson,rouetfeix}
which is the next order term in the expansion of the BBGKY hierarchy
in powers of $1/N$. For the spatially homogeneous HMF model, the
scaling of the relaxation time with $N$ is still controversial and
different scalings such as $t_R\sim N^2 t_D$
\cite{private} or $t_R\sim e^N t_D$ \cite{campa} have been
reported. An interesting problem
would be to determine numerically the scaling with $N$ of the
relaxation time of an axisymmetric distribution of point vortices. If
the collision term does not vanish at the next order of the $1/N$-expansion, this would imply a timescale scaling like $N^2 t_D$, but this
scaling has to be ascertained (this project is currently under way).

Note, finally, that the above results are only valid for axisymmetric
distributions of point vortices. Non-axisymmetric distributions are
described by a more complex kinetic equation (\ref{dev4}) which may
present new resonances allowing the system to reach the Boltzmann
distribution on a timescale of the order $t_R\sim N t_D$ (the natural
first order of the kinetic theory). A linear $N t_D$ scaling is indeed
observed numerically for the relaxation of a non-axisymmetric
distribution of point vortices \cite{kn}. However, very little is
known concerning the properties of Eq. (\ref{dev4}) and its
convergence (or not) towards the Boltzmann distribution. It could
approach the Boltzmann distribution (since entropy increases) without
reaching it exactly. New resonances also appear for spatially
inhomogeneous one dimensional systems with long-range interactions
\cite{angleaction,kindetail}. This may explain the linear $N t_D$
scaling of the relaxation time observed numerically for spatially
inhomogeneous one dimensional stellar systems
\cite{brucemiller,gouda,valageas,joyce} and for the spatially
inhomogeneous HMF model \cite{ruffoN}.  On the other hand, for the HMF
model, Yamaguchi {\it et al.} \cite{yamaguchi} find a relaxation time
scaling like $N^{\delta}t_D$ with $\delta\simeq 1.7$. In their
simulation, the initial distribution function is spatially homogeneous
but the collisional evolution makes it Vlasov unstable so that it
becomes spatially inhomogeneous. This corresponds to a dynamical phase
transition from a non-magnetized to a magnetized state as
theoretically studied in \cite{campachav}. In that case, the
relaxation time could be {\it intermediate} between $N^2 t_D$
(permanently homogeneous) and $N t_D$ (permanently
inhomogeneous). This argument (leading to $1<\delta<2$) may provide a
first step towards the explanation of the anomalous exponent
$\delta\simeq 1.7$ reported in \cite{yamaguchi}. The same phenomenon
(loss of Euler stability due to ``collisions'' and dynamical phase
transition from an axisymmetric distribution to a non-axisymmetric
distribution) could happen for the point vortex system.

\section{Relaxation of a test vortex in a bath: the Fokker-Planck equation}
\label{sec_tv}

In the previous sections, we have studied the evolution of the system
``as a whole''. In that approach, all the vortices are treated on the
same footing. We now consider the relaxation of a ``test'' vortex
(tagged particle) evolving in a steady distribution of ``field''
vortices.  If the field vortices are at statistical equilibrium,
described by the Boltzmann distribution (\ref{nv5}), their density
profile does not evolve at all. In that case, they form a thermal
bath. However, we shall also consider the case of an
out-of-equilibrium (i.e. non-Boltzmannian) bath corresponding to a
vorticity distribution $\omega(r)$ that is a stable steady state of
the 2D Euler equation with a monotonic profile of angular velocity. As
we have seen previously, this distribution does not change on a
timescale of order $Nt_D$. Since the relaxation time of a test vortex
in a bath is of order $(N/\ln N) t_D$ (see below) we can consider that
the distribution of the field vortices is ``frozen'' on this
timescale. They form therefore an out-of-equilibrium bath.

Let us call $P({\bf r},t)$ the probability density
of finding the test vortex at position ${\bf r}$ at time $t$. For
simplicity, we shall consider axisymmetric distributions (some results valid in the general case can be found in \cite{bbgky}). The evolution of $P({r},t)$
can be obtained from the kinetic equation (\ref{lb13}) by considering
that the distribution of the field vortices, described by the vorticity profile $\omega(r_2,t)$, is {\it fixed} \footnote{Indeed, we can interprete the kinetic equation (\ref{lb13}) as describing the ``collisions'' between a test vortex described by the variable $1$ and field vortices described by the running variable $2$. In Eq. (\ref{lb13}) all the vortices are equivalent so that the distribution of the field vortices $\omega(r_2,t)$ changes with time exactly like the distribution of the test vortex $\omega(r_1,t)$. In the bath approximation, we assume that the distribution of the field vortices $\omega(r_2)$ is prescribed.}. In the BBGKY hierarchy, this amounts to specializing a particular point vortex  in the system (the test vortex described by the variable $1$) and assuming that the other vortices (the field vortices described by the running variable $2$) are in a steady state. If we replace $\omega({r}_{1},t)$ by $P({r},t)$ and $\omega({r}_{2},t)$ by $\omega({r}')$, we get
\begin{eqnarray}
\frac{\partial P}{\partial t}(r,t)=2\pi^2 \gamma \frac{1}{r}
\frac{\partial}{\partial r} \int_0^{+\infty} r' dr' \chi(r,r',\Omega(r))
\delta(\Omega(r)-\Omega(r')) \left (\frac{1}{r}\frac{\partial}{\partial
r}-\frac{1}{r'}\frac{d}{d r'}\right )P(r,t)\omega(r'). \label{tv1}
\end{eqnarray}
This procedure transforms the integrodifferential equation (\ref{lb13}) describing the evolution of the system ``as a whole'' into a differential equation (\ref{tv1}) describing the evolution of a test vortex in a bath of field vortices.
Equation (\ref{tv1}) can be written in the form of a Fokker-Planck equation
\begin{equation}
\label{tv2}{\partial P\over\partial t}=\frac{1}{r}{\partial\over\partial r}\biggl\lbrack r\biggl (D{\partial P\over\partial r}-P V_r^{pol}\biggr )\biggr\rbrack,
\end{equation}
involving a diffusion coefficient
\begin{eqnarray}
D(r)=\frac{2\pi^{2}\gamma}{r^{2}}\int_{0}^{+\infty}r'dr'\chi(r,r',\Omega(r))\delta(\Omega(r)-\Omega(r'))\omega({r}'),
\label{tv3}
\end{eqnarray}
and a drift term due to the polarization
\begin{eqnarray}
V_{r}^{pol}(r)=\frac{2\pi^{2}\gamma}{r}\int_{0}^{+\infty}dr'\chi(r,r',\Omega(r))\delta(\Omega(r)-\Omega(r'))\frac{d\omega}{d r} ({r}').\label{tv4}
\end{eqnarray}
Physically, the diffusion coefficient is due to the fluctuations of the velocity
field produced by a discrete number of point vortices; it can be directly derived from the Kubo formula \cite{pre}.  On the other hand, the drift arises from the retroaction of the perturbation on the field vortices induced by the test vortex, just like in a polarization process; it can be directly derived from a linear response theory \cite{preR}. In the present case, the coefficients of
diffusion and drift depend on the position ${r}$ of the test
vortex.  Hence, it is more appropriate to write Eq. (\ref{tv1}) in a
form which is fully consistent with the general Fokker-Planck equation
\begin{equation}
\label{tv5} {\partial P\over\partial t}={1\over
2r}{\partial\over\partial r}\biggl\lbrack r{\partial\over\partial
r}\biggl  ({\langle (\Delta r)^{2}\rangle\over \Delta t}P\biggr
)\biggr\rbrack-{1\over r}{\partial\over\partial r}\biggl (rP{\langle
\Delta r\rangle\over \Delta t}\biggr ),
\end{equation}
with
\begin{equation}
\label{tv6}{\langle (\Delta r)^{2}\rangle\over 2 \Delta t}=D, \qquad
{\langle \Delta r\rangle\over \Delta t}\equiv V_r^{drift}={d D\over d
r}+V_r^{pol}.
\end{equation}
Substituting Eqs. (\ref{tv3}) and (\ref{tv4}) in Eq. (\ref{tv6}), and using an integration by parts, we find that the total drift term is given by
\begin{eqnarray}
V_r^{drift}(r)=2\pi^{2}\gamma\int_{0}^{+\infty} r r' dr' \omega(r') \left (\frac{1}{r}\frac{\partial}{\partial r}-\frac{1}{r'}\frac{d}{d r'}\right )\chi(r,r',\Omega(r))\delta(\Omega(r)-\Omega(r'))\frac{1}{r^2}.
\label{tv6b}
\end{eqnarray}
The two expressions (\ref{tv2}) and (\ref{tv5}) of the Fokker-Planck
equation have their own interest. The expression (\ref{tv5}) where the
diffusion coefficient is placed after the second derivative
$\partial^2(DP)$ involves the total drift ${\bf V}_{drift}$ and the
expression (\ref{tv2}) where the diffusion coefficient is placed
between the derivatives $\partial D\partial P$ isolates the part of
the drift ${\bf V}_{pol}$ due to the polarization. This expression is
directly connected to the form of the Lenard-Balescu-type equation
(\ref{lb13}). It has therefore a clear physical interpretation. The
Fokker-Planck equation (\ref{tv1}) and the expressions (\ref{tv3}),
(\ref{tv4}) and (\ref{tv6b}) of the diffusion coefficient and drift
term can also be obtained directly by calculating the first and second
moments of the increment of radial position of the test vortex using
the Klimontovich approach \cite{klim}.

If the profile of angular velocity of the field vortices $\Omega(r)$
is monotonic, we can use the identity
$\delta(\Omega(r)-\Omega(r'))=\delta(r-r')/|\Omega'(r)|$ and we find
that
\begin{eqnarray}
D(r)=2\pi^2\gamma
\frac{\chi(r,r,\Omega(r))}{|\Sigma(r)|}\omega(r),\label{tv7}
\end{eqnarray}
and
\begin{eqnarray}
V_{r}^{pol}(r)=2\pi^2\gamma
\frac{\chi(r,r,\Omega(r))}{|\Sigma(r)|}\frac{d\omega}{dr}(r),\label{tv8}
\end{eqnarray}
where $\Sigma(r)=r\Omega'(r)$ is the local shear. If we neglect collective effects, we can replace $\chi(r,r,\Omega(r))$ by
\begin{eqnarray}
\chi(r,r)=\sum_n |n|\hat{u}_n^2(r,r)=\frac{1}{8\pi^2}\sum_{n=1}^{+\infty}\frac{1}{n}\equiv \frac{1}{8\pi^2}\ln\Lambda,\label{tv8a}
\end{eqnarray}
where $\ln\Lambda\equiv \sum_{n=1}^{+\infty}\frac{1}{n}$ is a Coulomb
factor \footnote{For point vortices, the kinetic equation (\ref{lb13})
governing the evolution of the system as a whole does {\it not}
present any divergence contrary to the Lenard-Balescu equation in
plasma physics that presents a divergence at small scales.  However, a
logarithmic divergence at small scales occurs in the Fokker-Planck equation
(\ref{tv1}) when we make the bath approximation and consider that the
dominant contribution to the $\delta$-function comes from the
interaction at $r'=r$. This divergence comes from the assumption made
in the kinetic theory that the point vortices essentially follow the
streamlines produced by the mean flow (modified by collective
effects). This neglects the contribution of ``hard'' collisions with
small impact parameter that lead to more complex trajectories and
more complex interactions.  As explained in Sec. \ref{sec_bv}, these
hard collisions could be taken into account by keeping the
contribution of the fourth term in Eq. (\ref{bv6b}). Alternatively, we
can proceed heuristically and regularize the logarithmic divergence by
introducing a cut-off at the scale at which hard collisions come into
play \cite{preR,dubin2,cl}. With this regularization, it can be shown
that $\ln\Lambda\sim \frac{1}{2}\ln N$, leading to a $\ln N$ factor in
the Fokker-Planck equation (\ref{tv1}). Since only {\it distant}
collisions $r_2\neq r_1$ occur in the kinetic equation (\ref{lb13})
describing the evolution of the system as a whole, there is no
divergence at close separation, and therefore no $\ln N$ factor, in
that equation.}  that has to be regularized with appropriate cut-offs
as discussed in
\cite{preR,dubin2,cl}.  It is then found that $\ln\Lambda\sim
\frac{1}{2}\ln N$ in the thermodynamic limit $N\rightarrow +\infty$.
We note that the diffusion
coefficient and the drift due to the polarization are inversely proportional to the shear. Furthermore, the diffusion coefficient is proportional the vorticity profile of the field vortices while the drift is proportional to its gradient.   Comparing
Eqs. (\ref{tv7}) and (\ref{tv8}), we find that the drift velocity
is related to the diffusion coefficient by the relation
\begin{eqnarray}
V_{r}^{pol}(r)=D(r)\frac{d\ln\omega}{dr}(r). \label{tv9}
\end{eqnarray}
This can be viewed as a generalized form of Einstein relation for an
out-of-equilibrium distribution of field vortices. Combining the previous results we find that the
Fokker-Planck equation (\ref{tv2}) can be written
\begin{equation}
\label{tv10}{\partial P\over\partial
t}=\frac{1}{r}{\partial\over\partial r}\biggl\lbrack r D(r)\biggl
({\partial P\over\partial r}-P\frac{d\ln\omega}{dr}\biggr
)\biggr\rbrack,
\end{equation}
with a diffusion coefficient given by Eq. (\ref{tv7}).

If the field vortices are at statistical equilibrium (thermal bath),
their vorticity profile is the Boltzmann distribution
\begin{eqnarray}
\omega({r})=A\gamma e^{-\beta \gamma \psi_*(r)},
\label{tv11}
\end{eqnarray}
where $\psi_*(r)=\psi(r)+\frac{\Omega_{L}}{2}r^{2}$ is the relative stream function (see Sec. \ref{sec_nv}). We have
\begin{eqnarray}
\frac{d\omega}{dr}(r')=-\beta\gamma
\omega(r')\frac{d\psi_*}{dr}(r')=\beta\gamma\omega(r') (\Omega(r')-\Omega_{L})r',\label{tv12}
\end{eqnarray}
where we have used $\Omega(r)=-(1/r)d\psi/d r$.
Substituting this relation in Eq. (\ref{tv4}), using the
$\delta$-function to replace $\Omega(r')$ by $\Omega(r)$, using
$\Omega(r)-\Omega_{L}=-(1/r){d\psi_*}/{dr}$ and comparing the resulting
expression with Eq. (\ref{tv3}), we finally obtain
\begin{eqnarray}
V_{r}^{pol}(r)=-D(r)\beta\gamma \frac{d\psi_*}{dr}(r). \label{tv13}
\end{eqnarray}
The drift velocity  $ {\bf V}_{pol}=-D\beta\gamma\nabla\psi_*$ is perpendicular to the relative mean field velocity
$\langle {\bf V}_*\rangle=-{\bf z}\times\nabla\psi_*$. Furthermore, the
drift coefficient (or mobility) satisfies an Einstein relation
$\xi(r)=D(r)\beta\gamma$. We recall that the drift
coefficient and the diffusion coefficient depend on the position $r$
of the test vortex and that the temperature is negative in cases of
physical interest. We also stress that the Einstein relation is valid for
the drift ${\bf V}_{pol}$ due to the polarization, not for the total
drift ${\bf V}_{drift}$ which has a more complicated expression. We do not have this subtlety for the usual
Brownian motion where the diffusion coefficient is constant.
For a thermal bath, using Eq. (\ref{tv13}), the Fokker-Planck equation
(\ref{tv2}) can be written
\begin{equation}
\label{tv14}{\partial P\over\partial
t}=\frac{1}{r}{\partial\over\partial r}\biggl\lbrack r D(r)\biggl
({\partial P\over\partial r}+\beta\gamma P\frac{d\psi_*}{dr}\biggr
)\biggr\rbrack,
\end{equation}
where $D(r)$ is given by Eq. (\ref{tv3}) with Eq. (\ref{tv11}). Of
course, if the profile of angular velocity of the Boltzmann
distribution is monotonic, we find that Eq. (\ref{tv10}) with
Eq. (\ref{tv11}) returns Eq. (\ref{tv14}) with a diffusion coefficient
given by Eq. (\ref{tv7}) with Eq. (\ref{tv11}). Finally, we note that
the systematic drift $ {\bf V}_{pol}=-D\beta\gamma\nabla\psi_*$ of a
point vortex \cite{preR} is the counterpart of the dynamical friction
${\bf F}_{pol}=-D_{\|}\beta m {\bf v}$ of a star
\cite{chandrafriction}.  Similarly, the Smoluchowski-type form of the
Fokker-Planck equation (\ref{tv14}) describing the relaxation a point
vortex in a ``sea'' of field vortices \cite{preR} is the counterpart
of the Kramers-type form of the Fokker-Planck equation describing the
relaxation of a star in a globular cluster
\cite{chandrafriction}. This is an aspect of the numerous analogies
that exist between two-dimensional vortices and stellar systems
\cite{houchesPH}.

The Fokker-Planck equations (\ref{tv10}) and (\ref{tv14}) have been
studied for different types
of bath distribution in \cite{cl}.  The distribution of the test vortex $P({\bf
r},t)$ relaxes towards the distribution  of the bath $\omega ({\bf r})/\Gamma$
on a typical timescale
\begin{equation}
\label{tv15}
t_{R}^{bath}\sim \frac{N}{\ln N}t_D,
\end{equation}
where the logarithmic correction comes from the scaling with $N$ of
the Coulombian factor $\ln\Lambda$.  However, the relaxation process
towards the steady state is very peculiar and differs from the usual
exponential relaxation of Brownian particles. In particular, the
evolution of the front profile in the tail of the distribution is very
slow, scaling like $(\ln t)^{1/2}$, and the temporal correlation
function $\langle r(0)r(t)\rangle$ decreases algebraically, like $\ln
t/t$ (for a thermal bath), instead of exponentially. This is due to
the rapid decay of the diffusion coefficient $D(r)$ for large
$r$. Similar results had been found earlier for the HMF model
\cite{bd,cl1}.

Finally, we stress that the evolution of the system ``as a whole'' is very different from the evolution of a test vortex in a bath. We have seen in Sec. \ref{sec_num} that the relaxation time of the system as a whole is strictly larger than $Nt_D$ (for axisymmetric distributions) while the relaxation time of a test vortex in a bath is of the order $(N/\ln N)t_D$. In particular, a steady state of the 2D Euler equation with a monotonic profile of angular velocity does not change on this timescale. This justifies our procedure of developing a bath approximation for out-of-equilibrium (i.e. non-Boltzmannian) distributions of the field vortices.

\section{Conclusion}
\label{sec_conclusion}

The statistical mechanics of two-dimensional point vortices was first
considered by Onsager \cite{onsager} in a seminal paper.  The point
vortex gas is a $N$-body Hamiltonian system with long-range
interactions that shares many analogies with electric charges in a
plasma and stars in a stellar system. Like stellar systems, the point
vortex gas displays two successive types of relaxation: A fast, or
violent, relaxation due to {\it mean field} effects followed by a slow
relaxation due to {\it discrete} effects.

In a first regime, before the correlations (``collisions'') due to
graininess and finite $N$ effects  comes into
play, the evolution of the smooth vorticity field is governed by the
2D Euler equation. In practice, the validity of the 2D Euler equation
is huge since the collisional relaxation time diverges with the number
of point vortices. Starting from a generically unstable or unsteady
initial condition, the 2D Euler-Poisson system experiences a process
of violent relaxation leading to a non-Boltzmannian quasi-stationary
state (QSS) which is a steady state of the 2D Euler equation on the
coarse-grained scale.  These QSSs correspond to large-scale vortices
observed in 2D turbulence like monopoles, dipoles, tripoles...  One
can attempt to predict these QSSs in terms of statistical mechanics by
using the Miller-Robert-Sommeria
\cite{miller,rs} theory which is the hydrodynamical version of the
Lynden-Bell \cite{lb} theory. This theory is, however, based on an
assumption of ergodicity (or efficient mixing) which may not always be
fulfilled. This leads to the complicated problem of {\it incomplete
relaxation} which restricts the range of validity of the MRS (or
Lynden-Bell) statistical theory. For example, ``vortex crystals''
\cite{fine} could correspond to metastable states that are only
partially mixed.

On a longer timescale, ``distant collisions'' (i.e. correlations,
graininess or finite $N$ effects) between vortices must be taken into
account.  At the order $1/N$, this collisional evolution is described
by the kinetic equation (\ref{dev4}) which reduces to the more
explicit form (\ref{lb13}) for an axisymmetric distribution of point
vortices. However, this kinetic equation does {\it not} relax towards
the Boltzmann distribution predicted by statistical mechanics for
$t\rightarrow +\infty$.  Indeed, the evolution stops when the profile
of angular velocity becomes monotonic. The relaxation towards
statistical equilibrium may occur on a very long time, larger than $N
t_D$. To settle this issue, we have to develop the kinetic theory at
the order $1/N^2$, $1/N^3$,... by taking into account three-body,
four-body,...  correlation functions. It is also possible that the
point vortex gas never reaches the Boltzmann distribution: Mixing by
``collisions'' may not be efficient enough. For non-axisymmetric
distributions of point vortices, the relaxation time may be reduced,
due to the occurrence of additional resonances, and achieve the
natural timescale $Nt_D$ predicted by the first order kinetic theory
\cite{kindetail}. This linear scaling has been observed in numerical
simulations \cite{kn}. It is, however, hard to prove that the general
kinetic equation (\ref{dev4}) will reach the Boltzmann
distribution. It may approach it without reaching it exactly.

Finally, the stochastic motion of a test vortex in a ``sea''
(bath) of field vortices can be described in terms of a Fokker-Planck
equation involving a diffusion term and a drift term. For a thermal
bath, they are connected to each other by an Einstein relation. The
diffusion coefficient (resp. drift term) is proportional to the
vorticity distribution (resp. to the gradient of the vorticity
distribution) of the field vortices and inversely proportional to the
local shear.  The distribution of the test vortex relaxes towards the
distribution of the field vortices on a timescale of the order $(N/\ln
N)t_{D}$ but the relaxation process is peculiar \cite{cl}.

For the point vortex gas, it is important to clearly distinguish the
phase of violent collisionless (correlationless) relaxation towards a
non-Boltzmannian QSS, taking place on a few dynamical times $t_D$,
from the slow collisional (correlational) relaxation towards the
Boltzmann distribution taking place on a very long timescale
$>Nt_D$. In this sense, the MRS theory and the Boltzmann theory
describe two completely different regimes. This distinction between
collisional (Boltzmann) and collisionless (Lynden-Bell) relaxation has
been clearly understood for a long time in astrophysics \cite{henon,bt} but
it may be less well-known in the context of two-dimensional point
vortex dynamics.  This is because, implicitly, the Onsager theory has
been used to describe the equilibrium state resulting from the phase
of violent relaxation before the Lynden-Bell and the MRS theory were
formulated. In fact, the Boltzmann distribution resulting from the
Onsager theory has two completely different interpretations: (i) it
can be viewed as an {\it approximation} of the MRS distribution
arising from the collisionless mixing of the 2D Euler equation on a
``short'' timescale; (ii) it can be viewed as the ordinary Boltzmann
equilibrium state arising from the collisional evolution of the point
vortex gas on a long timescale. In the first case, mixing is due to a
purely mean field process  while, in the second
case, mixing is due to discrete interactions between vortices
(collisions). These two regimes correspond to a different ordering of the 
$N\rightarrow +\infty$ and $t\rightarrow +\infty$ limits.

We must also clearly distinguish the evolution of the system of point
vortices ``as a whole'' from the relaxation of a test vortex in a
bath. They correspond to completely different situations and they are
described by different equations. The evolution of the system as a
whole is described by the integrodifferential equation (\ref{lb13})
similar to the Lenard-Balescu equation in plasma physics, while the
relaxation of a test vortex in a bath is described by a differential
equation (\ref{tv1}) similar to the Fokker-Planck equation in Brownian
theory. 

Finally, we may compare the evolution of a large, but finite, number
of point vortices with the evolution of a real continuous vorticity
field. On a ``short'' timescale, we can neglect correlations and
finite $N$ effects in the evolution of point vortices and we can
neglect viscosity in the evolution of the real continuous vorticity
field. In that case, the two systems can be approximated by the 2D
Euler equation and they should behave similarly. In particular, they
may experience a process of violent relaxation towards a QSS. On a
``long'' timescale, the evolution of the point vortex gas and of the
real continuous vorticity field will differ. Point vortices will feel
correlations due to finite $N$ effects and will (with the reserve that
we have given in this paper) relax towards the Boltzmann distribution
while a continuous vorticity field will feel inherent viscosity and
will decay to zero.

\appendix

\section{The linearized 2D Euler equation}
\label{sec_euler}

The 2D Euler equation can be written
\begin{eqnarray}
\label{euler1}
\frac{\partial \omega}{\partial t}+{\bf u}\cdot \nabla\omega=0,
\end{eqnarray}
where ${\bf u}=-{\bf z}\times \nabla\psi$ is the velocity field (${\bf
z}$ is a unit vector normal to the plane of the flow). The stream
function $\psi({\bf r},t)$ is related to the vorticity $\omega({\bf
r},t)$ by the Poisson equation $\Delta\psi=-\omega$. The potential of
interaction $u({\bf r},{\bf r}')$, which is the Green function of the
Laplacian operator $\Delta$, is defined by $\Delta u({\bf r},{\bf
r}')=-\delta({\bf r}-{\bf r}')$. In an infinite domain, $u=u(|{\bf
r}-{\bf r}'|)$ only depends on the absolute distance between two
points and is given by $u(|{\bf r}-{\bf r}'|)=-(1/2\pi)\ln|{\bf
r}-{\bf r}'|$.  This corresponds to a Newtonian (or Coulombian)
interaction in two dimensions. Therefore, the stream function is
related to the vorticity field by an expression of the form $\psi({\bf
r},t)=\int u(|{\bf r}-{\bf r}'|)\omega({\bf r'},t)\, d{\bf r}$. This
can be written as a product of convolution $\psi=u*\omega$.

Let us consider a small perturbation $\delta\omega({\bf r},t)$ around a steady state $\omega({\bf r})$ of the 2D Euler equation. We write $\omega({\bf r},t)=\omega({\bf r})+\delta\omega({\bf r},t)$, $\psi({\bf r},t)=\psi({\bf r})+\delta\psi({\bf r},t)$ and ${\bf u}({\bf r},t)={\bf u}({\bf r})+\delta {\bf u}({\bf r},t)$. Substituting these decompositions in Eq. (\ref{euler1}) and neglecting the quadratic terms, we obtain the linearized Euler equation
\begin{equation}
\frac{\partial\delta\omega}{\partial t}+{\bf u}\cdot\nabla\delta\omega+\delta{\bf u}\cdot\nabla\omega=0.
\label{euler2}
\end{equation}
If we restrict ourselves to axisymmetric mean flows, then  ${\bf u}({\bf r})=u(r){\bf e}_{\theta}$ with $u(r)=-\frac{\partial\psi}{\partial r}(r)=\Omega(r)r$, where  $\Omega(r)=\frac{1}{r^{2}}\int_0^r \omega(r')r'\, dr'$  is the angular velocity (see Sec. 6.1 of \cite{cl}). In that case, Eq. (\ref{euler2}) becomes
\begin{equation}
\frac{\partial\delta\omega}{\partial t}+\Omega(r)\frac{\partial\delta\omega}{\partial\theta}+\frac{1}{r}\frac{\partial\delta\psi}{\partial \theta}\frac{\partial\omega}{\partial r}=0.
\label{euler3}
\end{equation}
To solve the initial value problem, it is convenient to introduce Fourier-Laplace transforms. The Fourier-Laplace transform of the perturbation of the vorticity field $\delta\omega$ is defined by
\begin{equation}
\delta \tilde\omega(n,r,\sigma)=\int_{0}^{2\pi}\frac{d\theta}{2\pi}\int_{0}^{+\infty}dt\, e^{-i(n\theta-\sigma t)}\delta\omega(\theta,r,t).
\label{euler4}
\end{equation}
This expression for the Laplace transform is valid for ${\rm Im}(\sigma)$ sufficiently large. For the remaining part of the complex $\sigma$ plane, it is defined by an analytic continuation. The inverse transform is
\begin{equation}
\delta \omega(\theta,r,t)=\sum_{n=-\infty}^{+\infty}\int_{\cal C}\frac{d\sigma}{2\pi}\, e^{i(n\theta-\sigma t)}\delta\tilde\omega(n,r,\sigma),
\label{euler5}
\end{equation}
where the Laplace contour ${\cal C}$ in the complex $\sigma$ plane must pass above all poles of the integrand. Similar expressions hold for the perturbation of the stream function $\delta\psi$. If we take the  Fourier-Laplace transform of the linearized Euler equation (\ref{euler3}), we obtain
\begin{equation}
-\delta\hat{\omega}(n,r,0)-i\sigma\delta\tilde\omega(n,r,\sigma)+in\Omega(r)\delta\tilde\omega(n,r,\sigma)+in\frac{1}{r}\frac{\partial\omega}{\partial r}\delta\tilde\psi(n,r,\sigma)=0,
\label{euler6}
\end{equation}
where the first term is the spatial Fourier transform of the initial value
\begin{equation}
\delta\hat\omega(n,r,0)=\int_{0}^{2\pi}\frac{d\theta}{2\pi}\, e^{-in\theta}\delta\omega(\theta,r,0).
\label{euler7}
\end{equation}
The foregoing equation can be rewritten
\begin{equation}
\delta\tilde\omega (n,r,\sigma)=\frac{n\frac{1}{r}\frac{\partial\omega}{\partial r}}{\sigma-n\Omega(r)}\delta\tilde\psi(n,r,\sigma)-\frac{\delta\hat\omega(n,r,0)}{i(\sigma-n\Omega(r))},
\label{euler8}
\end{equation}
where the first term on the right hand side corresponds to ``collective effects'' and the second term is related to the initial condition. The perturbation of the stream function is related to the perturbation of the vorticity by the Poisson equation
\begin{equation}
\frac{1}{r}\frac{\partial}{\partial r} r \frac{\partial \delta\psi}{\partial r}+\frac{1}{r^2}\frac{\partial^2\delta\psi}{\partial\theta^2}=-\delta\omega.
\label{euler9}
\end{equation}
Taking the Fourier-Laplace transform of this equation,  we obtain
\begin{equation}
\left\lbrack \frac{1}{r}\frac{\partial}{\partial r} r \frac{\partial}{\partial r}-\frac{n^2}{r^{2}}\right \rbrack \delta\tilde\psi(n,r,\sigma)=-\delta\tilde\omega(n,r,\sigma).
\label{euler10}
\end{equation}
Substituting Eq. (\ref{euler8}) in Eq. (\ref{euler10}), we find that
\begin{equation}
\left\lbrack \frac{1}{r}\frac{\partial}{\partial r} r \frac{\partial}{\partial r}-\frac{n^2}{r^{2}}+\frac{n\frac{1}{r}\frac{\partial\omega}{\partial r}}{\sigma-n\Omega(r)}\right \rbrack \delta\tilde\psi(n,r,\sigma)=\frac{\delta\hat\omega(n,r,0)}{i(\sigma-n\Omega(r))}.
\label{euler11}
\end{equation}
Therefore, the Fourier-Laplace transform of the perturbation of the stream function is related to the initial condition by
\begin{equation}
\delta\tilde\psi(n,r,\sigma)=-\int_{0}^{+\infty}2\pi r'\, dr'\, G(n,r,r',\sigma)\frac{\delta\hat\omega(n,r',0)}{i(\sigma-n\Omega(r'))},
\label{euler12}
\end{equation}
where the Green function $G(n,r,r',\sigma)$ is defined by
\begin{equation}
\left\lbrack \frac{1}{r}\frac{\partial}{\partial r} r \frac{\partial}{\partial r}-\frac{n^2}{r^{2}}+\frac{n\frac{1}{r}\frac{\partial\omega}{\partial r}}{\sigma-n\Omega(r)}\right \rbrack G(n,r,r',\sigma)=-\frac{\delta(r-r')}{2\pi r}.
\label{euler13}
\end{equation}
If we neglect collective effects in the  foregoing equation, we obtain
\begin{equation}
\left\lbrack \frac{1}{r}\frac{\partial}{\partial r} r \frac{\partial}{\partial r}-\frac{n^2}{r^{2}}\right \rbrack G_{bare}(n,r,r')=-\frac{\delta(r-r')}{2\pi r}.
\label{euler14}
\end{equation}
Therefore,  $G_{bare}(n,r,r')=\hat{u}_n(r,r')$ is the Fourier transform of the ``bare'' potential of interaction $u$ that is solution of the Poisson equation $\Delta u=-\delta$, while $G(n,r,r',\sigma)$ is the Laplace-Fourier transform of the   potential of interaction ``dressed'' by its polarization cloud, i.e. taking collective effects into account.

Using the previous formulae, we can now relate the Laplace-Fourier transform $\delta\tilde\omega (n,r,\sigma)$ of the perturbation of the vorticity to the initial condition $\delta\hat\omega(n,r,0)$. Substituting Eq. (\ref{euler12}) in Eq. (\ref{euler8}), we obtain
\begin{equation}
\delta\tilde\omega (n,r,\sigma)=-\frac{n\frac{1}{r}\frac{\partial\omega}{\partial r}}{\sigma-n\Omega(r)}
\int_{0}^{+\infty}2\pi r''\, dr''\, G(n,r,r'',\sigma)\frac{\delta\hat\omega(n,r'',0)}{i(\sigma-n\Omega(r''))}
-\frac{\delta\hat\omega(n,r,0)}{i(\sigma-n\Omega(r))}.
\label{euler15}
\end{equation}
If we consider an initial condition of the form $\delta\omega(r,\theta,0)=\gamma \delta(\theta-\theta')\delta(r-r')/r$, implying
\begin{equation}
\delta\hat\omega(n,r,0)=\frac{\gamma}{2\pi r}e^{-in\theta'}\delta(r-r'),
\label{euler16}
\end{equation}
we find that
\begin{equation}
\delta\tilde\omega (n,r,\sigma)=-\frac{n\frac{1}{r}\frac{\partial\omega}{\partial r}}{\sigma-n\Omega(r)}
\gamma  e^{-in\theta'} \frac{G(n,r,r',\sigma)}{i(\sigma-n\Omega(r'))}
-\gamma\frac{e^{-in\theta'}}{2\pi r}\frac{\delta(r-r')}{i(\sigma-n\Omega(r))}.
\label{euler17}
\end{equation}
As shown in Sec. \ref{sec_axi}, this equation gives the expression of the propagator $U_n(r,r',\sigma)$ [see Eq. (\ref{axi10})]. This operator is also called the resolvent operator as it connects $\delta\tilde\omega (n,r,\sigma)$ to its initial value:
\begin{equation}
\delta\tilde\omega (n,r,\sigma)=\int_0^{+\infty} U_n(r,r'',\sigma) \delta\hat{\omega}(n,r'',0)\, 2\pi r'' \, dr''.
\label{euler18}
\end{equation}

\section{An integral equation}
\label{sec_int}

The perturbation of the stream function is related to the perturbation of the vorticity by an expression of the form
\begin{equation}
\delta\psi({\bf r},t)=\int u(|{\bf r}-{\bf r}'|)\delta\omega({\bf r'},t)\, d{\bf r}.
\label{int1}
\end{equation}
The potential of interaction can be written
\begin{equation}
u(|{\bf r}-{\bf r}'|)=u\left (\sqrt{r^2+r'^2-2rr'\cos(\theta-\theta')}\right )=u(r,r',\phi),
\label{int2}
\end{equation}
where $\phi=\theta-\theta'$. We note that $u(r,r'\phi)=u(r',r,\phi)$ and $u(r,r',-\phi)=u(r,r',\phi)$. Due to its $\phi$-periodicity, the potential of interaction can be decomposed in Fourier series as
\begin{equation}
u(r,r',\phi)=\sum_n e^{i n \phi}\hat{u}_n(r,r'),\qquad \hat{u}_n(r,r')=\int_{0}^{2\pi}\frac{d\phi}{2\pi}u(r,r',\phi)\cos(n\phi).
\label{int3}
\end{equation}
For the Coulombian potential  $u(|{\bf r}-{\bf r}'|)=-(1/2\pi)\ln|{\bf r}-{\bf r}'|$, satisfying $u(r,r',\phi)=-(1/4\pi)\ln(r^2+r'^2-2rr'\cos\phi)$, the integrals in Eq. (\ref{int3}) can be performed analytically \cite{kindetail}. We find that
\begin{equation}
\hat{u}_n(r,r')=\frac{1}{4\pi |n|}\left (\frac{r_<}{r_>}\right )^{|n|},\qquad \hat{u}_0(r,r')=-\frac{1}{2\pi}\ln r_>.
\label{int3a}
\end{equation}
Therefore, the potential of interaction can be written
\begin{equation}
u(r,r',\phi)=-\frac{1}{2\pi}\ln r_>+\frac{1}{4\pi}\sum_{n\neq 0}\frac{1}{|n|}\left (\frac{r_<}{r_>}\right )^{|n|}e^{in\phi},
\label{int3b}
\end{equation}
which is just the Fourier decomposition of the logarithm in two dimensions. Taking the Fourier-Laplace transform of Eq. (\ref{int1}) and using the fact that the integral is a product of convolution, we obtain
\begin{equation}
\delta\tilde\psi(n,r,\sigma)=2\pi \int_0^{+\infty} r'dr'\,  \hat{u}_n(r,r') \delta\tilde\omega(n,r',\sigma).
\label{int4}
\end{equation}
If we substitute Eq. (\ref{euler8}) in Eq. (\ref{int4}), we obtain the equation
\begin{equation}
\delta\tilde\psi(n,r,\sigma)-2\pi \int_0^{+\infty} r'dr'\,  \hat{u}_n(r,r')\frac{n\frac{1}{r'}\frac{\partial\omega}{\partial r'}(r')}{\sigma-n\Omega(r')}\delta\tilde\psi(n,r',\sigma) =-2\pi \int_0^{+\infty} r'dr'\,  \hat{u}_n(r,r') \frac{\delta\hat\omega(n,r',0)}{i(\sigma-n\Omega(r'))},
\label{int5}
\end{equation}
which is equivalent to Eq. (\ref{euler12}). This implies that the Green function $G(n,r,r',\sigma)$ satisfies an integral equation of the form
\begin{equation}
G(n,r,r',\sigma)-2\pi \int_0^{+\infty} r''dr''\,  \hat{u}_n(r,r'')\frac{n\frac{1}{r''}\frac{\partial\omega}{\partial r''}(r'')}{\sigma-n\Omega(r'')}G(n,r'',r',\sigma) =\hat{u}_n(r,r'),
\label{int6}
\end{equation}
which is equivalent to Eq. (\ref{euler13}). If we neglect collective effects, Eq. (\ref{int5}) reduces to
\begin{equation}
\delta\tilde\psi(n,r,\sigma) =-2\pi \int_0^{+\infty} r'dr'\,  \hat{u}_n(r,r') \frac{\delta\hat\omega(n,r,0)}{i(\sigma-n\Omega)}.
\label{int7}
\end{equation}
Comparing Eq. (\ref{int7}) with Eq. (\ref{euler12}), we see that the bare Green function $G_{bare}(n,r,r')=\hat{u}_n(r,r')$ is the Fourier transform of the potential of interaction $u$.

\section{General kinetic equation without collective effects}
\label{sec_nocoll}

If we neglect collective effects, the first two equations (\ref{bv7}) and (\ref{bv8}) of the BBGKY-like hierarchy reduce to
\begin{eqnarray}
\frac{\partial\omega}{\partial t}({\bf r}_1,t)+\frac{N-1}{N}\langle {\bf V}\rangle({\bf r}_1,t)\cdot \frac{\partial \omega}{\partial {\bf r}_{1}}({\bf r}_1,t)=-\gamma
\frac{\partial}{\partial {\bf r}_{1}}\cdot \int {\bf V}(2\rightarrow 1)g({\bf r}_{1},{\bf r}_{2},t)\, d{\bf r}_{2},
\label{nocoll1}
\end{eqnarray}
\begin{eqnarray}
\label{nocoll2} \frac{\partial g}{\partial t}({\bf r}_1,{\bf r}_2,t)+\left\lbrack \langle {\bf V}\rangle({\bf r}_1,t)\cdot \frac{\partial}{\partial {\bf r}_{1}}+\langle{\bf V}\rangle({\bf r}_2,t)\cdot \frac{\partial}{\partial {\bf r}_{2}}\right\rbrack g({\bf r}_1,{\bf r}_2,t)\nonumber\\
=-\frac{1}{\gamma^{2}}\left\lbrack
 \tilde{\bf V}(2\rightarrow 1)\cdot \frac{\partial}{\partial {\bf r}_1}+\tilde{\bf V}(1\rightarrow 2)\cdot \frac{\partial}{\partial {\bf r}_2}\right\rbrack \omega({\bf r}_1,t)\omega({\bf r}_2,t).
\end{eqnarray}
Equation (\ref{nocoll2}) is just a first order differential equation in time. It can be written symbolically as
\begin{eqnarray}
\label{nocoll2b} \frac{\partial g}{\partial t}+{\cal L}g=S\lbrack \omega\rbrack,
\end{eqnarray}
where ${\cal L}$ is an advection operator and $S$ is a source term: The correlation function is transported by the mean flow (l.h.s.) and modified by two-body collisions (r.h.s.). This equation can be solved by the method of characteristics. Introducing the Green function
\begin{eqnarray}
\label{nocoll2c}
G(t,t')={\rm exp}\left\lbrace -\int_t^{t'}{\cal L}(\tau)\, d\tau\right\rbrace,
\end{eqnarray}
constructed
with the smooth velocity field $\langle {\bf V}\rangle$, we obtain
\begin{eqnarray}
\label{nocoll2d} g({\bf r}_1,{\bf r}_2,t)=-\frac{1}{\gamma^{2}}\int_0^t d\tau\, G(t,t-\tau) \left\lbrack
 \tilde{\bf V}(2\rightarrow 1)\cdot \frac{\partial}{\partial {\bf r}_1}+\tilde{\bf V}(1\rightarrow 2)\cdot \frac{\partial}{\partial {\bf r}_2}\right\rbrack \omega({\bf r}_1,t-\tau)\omega({\bf r}_2,t-\tau),
\end{eqnarray}
where we have assumed that no correlation is present initially so that $g({\bf r}_1,{\bf r}_2,t=0)=0$ (if correlations are present initially, it can be shown that they are rapidly washed out). Substituting this result in Eq. (\ref{nocoll1}), we obtain
\begin{eqnarray}
\frac{\partial \omega}{\partial t}({\bf r}_1,t)+\frac{N-1}{N}\langle {\bf
V}\rangle({\bf r}_1,t)\cdot {\partial \omega\over \partial {\bf r}_{1}}({\bf r}_1,t)
=\frac{\partial}{\partial {r}_1^{\mu}}\int_0^t d\tau \int d{\bf
r}_{2} {V}^{\mu}(2\rightarrow
1)G(t,t-\tau)\nonumber\\
\times  \left \lbrack {\tilde{V}}^{\nu}(2\rightarrow 1)
{\partial\over\partial { r}_{1}^{\nu}}+{\tilde{V}}^{\nu}(1\rightarrow
2) {\partial\over\partial {r}_{2}^{\nu}}\right
\rbrack\omega({\bf
r}_1,t-\tau)\frac{\omega}{\gamma}({\bf r}_2,t-\tau). \label{nocoll3}
\end{eqnarray}
In writing this equation, we have adopted a Lagrangian point of view:
the coordinates ${\bf r}_i$ following the Greenian must be viewed as
${\bf r}_i(t-\tau)={\bf r}_i(t)-\int_0^{\tau}ds\, \langle {\bf
V}\rangle({\bf r}_i(t-s),t-s)\, ds$. The generalized Landau equation
(\ref{nocoll3}) which is valid for possibly non-axisymmetric flows and
which takes into account non-Markovian effects has been derived in our
previous papers from the projection operator formalism \cite{pre}, the
BBGKY hierarchy \cite{bbgky} and the quasilinear theory based on the
Klimontovich equation \cite{bbgky}. The generalized Lenard-Balescu
equation (\ref{dev3}) can be viewed as an extension of this equation
taking collective effects into account. Note that the Markovian
approximation may not be justified in every situation since it has
been found numerically that point vortices can exhibit long jumps
(L\'evy flights) and strong correlations \cite{kawahara,yoshida}.

If we implement the Bogoliubov ansatz and extend the integration on time $\tau$ to infinity, we get
\begin{eqnarray}
\frac{\partial \omega}{\partial t}({\bf r}_1,t)+\frac{N-1}{N}\langle {\bf
V}\rangle({\bf r}_1,t)\cdot {\partial \omega\over \partial {\bf r}_{1}}({\bf r}_1,t)
=\frac{\partial}{\partial {r}_1^{\mu}}\int_0^{+\infty} d\tau \int d{\bf
r}_{2} {V}^{\mu}(2\rightarrow
1,t)G(t,t-\tau)\nonumber\\
\times  \left \lbrack {\tilde{V}}^{\nu}(2\rightarrow 1)
{\partial\over\partial { r}_{1}^{\nu}}+{\tilde{V}}^{\nu}(1\rightarrow
2) {\partial\over\partial {r}_{2}^{\nu}}\right
\rbrack\omega({\bf
r}_1,t)\frac{\omega}{\gamma}({\bf r}_2,t),\label{nocoll4}
\end{eqnarray}
where, now, ${\bf r}_i(t-\tau)={\bf r}_i(t)-\int_0^{\tau}ds\, \langle {\bf V}\rangle({\bf r}_i(t-s),t)\, ds$. If we consider axisymmetric flows, this equation can be simplified \cite{pre,bbgky,kindetail} and we obtain the Landau-type equation (\ref{lan6}). If we take collective effects into account, Eq. (\ref{nocoll4}) is replaced by Eq. (\ref{dev4}) which reduces to the Lenard-Balescu-type kinetic equation (\ref{lb12}) for axisymmetric flows.

If we assume right from the beginning that the mean flow is axisymmetric, Eqs. (\ref{nocoll1}) and (\ref{nocoll2}) can be rewritten as
\begin{eqnarray}
\frac{\partial\omega}{\partial t}(r_1,t)=-\gamma^2\frac{1}{r_1}\frac{\partial}{\partial r_1}\int_0^{+\infty} r_2\, dr_2\int_{0}^{2\pi} d\theta_2\, \frac{\partial u}{\partial \theta_1}(r_1,r_2,\theta_1-\theta_2) g(r_1,r_2,\theta_1-\theta_2,t),
\label{nocoll5}
\end{eqnarray}
\begin{eqnarray}
\label{nocoll6} \frac{\partial g}{\partial t}({r}_1,{r}_2,\theta_1-\theta_2,t)+\left\lbrack\Omega(r_1,t)-\Omega(r_2,t)\right\rbrack\frac{\partial g}{\partial \theta_1}({r}_1,{r}_2,\theta_1-\theta_2,t)\nonumber\\
=-\frac{\partial u}{\partial\theta_1}(r_1,r_2,\theta_1-\theta_2)\left (\frac{1}{r_1}\frac{\partial}{\partial {r}_1}-\frac{1}{r_2}\frac{\partial}{\partial {r}_2}\right )\omega({r}_1,t)\omega({r}_2,t).
\end{eqnarray}
Introducing the Fourier transforms of the potential of interaction and of the correlation function, we obtain
\begin{eqnarray}
\label{nocoll7} \frac{\partial\omega}{\partial t}(r_1,t)=2i\pi\gamma^2\frac{1}{r_1}\frac{\partial}{\partial r_1}\int_0^{+\infty}r_2\, dr_2\sum_n  n \, \hat{u}_n(r_1,r_2)\hat{g}_n(r_1,r_2,t),
\end{eqnarray}
\begin{eqnarray}
\label{nocoll8} \frac{d\hat{g}_n}{dt}({r}_1,{r}_2,t)+in\left\lbrack\Omega(r_1,t)-\Omega(r_2,t)\right\rbrack \hat{g}_n=-\frac{i}{\gamma}  n \, \hat{u}_n(r_1,r_2)\left (\frac{1}{r_1}\frac{\partial}{\partial {r}_1}-\frac{1}{r_2}\frac{\partial}{\partial {r}_2}\right )\omega({r}_1,t)\omega({r}_2,t).
\end{eqnarray}
With the Bogoliubov ansatz, Eq. (\ref{nocoll7}) becomes
Eq. (\ref{axi5}). On the other hand, the asymptotic value of the
correlation function can be obtained by taking the Laplace transform
of Eq. (\ref{nocoll8}) and considering the limit $t\rightarrow
+\infty$. Equivalently, it is obtained by making the substitution
\begin{eqnarray}
\label{nocoll9} \frac{d\hat{g}_n}{dt}({r}_1,{r}_2,t)\rightarrow \lim_{\epsilon\rightarrow 0^+} \epsilon \hat{g}_n({r}_1,{r}_2,+\infty),
\end{eqnarray}
in Eq. (\ref{nocoll8}). This yields
\begin{eqnarray}
\label{nocoll10} \hat{g}_n({r}_1,{r}_2,+\infty)=-\frac{1}{\gamma}  \frac{n \, \hat{u}_n(r_1,r_2)}{n\left\lbrack\Omega(r_1)-\Omega(r_2)\right\rbrack-i0^+}\left (\frac{1}{r_1}\frac{\partial}{\partial {r}_1}-\frac{1}{r_2}\frac{\partial}{\partial {r}_2}\right )\omega({r}_1)\omega({r}_2).
\end{eqnarray}
Substituting Eq. (\ref{nocoll10}) in Eq. (\ref{axi5}), we obtain Eq. (\ref{lan3}) which finally leads to the Landau-type equation (\ref{lan6}).

%\begin{acknowledgements}
%\end{acknowledgements}

% BibTeX users please use
%\bibliographystyle{spmpsci}


\end{document}